\documentclass{article}
\usepackage{ijcai24}

\usepackage{times}
\usepackage{soul}
\usepackage{url}
\usepackage[hidelinks]{hyperref}
\usepackage[utf8]{inputenc}
\usepackage[small]{caption}
\usepackage{graphicx}
\usepackage{amsmath}
\usepackage{amsthm}
\usepackage{natbib}
\usepackage{booktabs}
\usepackage[switch]{lineno}

\usepackage{bm}
\usepackage{amsfonts}
\usepackage{xspace}
\usepackage{adjustbox}
\usepackage{xcolor}
\usepackage{algorithm}
\usepackage{algorithmic}
\usepackage{fancyhdr}
\usepackage{mathrsfs}
\usepackage{amsthm}
\usepackage{amssymb}

\newcommand{\argmax}{\operatornamewithlimits{argmax}}
\newcommand{\term}[1]{\textbf{\textit{#1}}}

\newcommand{\cA}{\mathcal{A}}

\newcommand{\cG}{\mathcal{G}}

\newcommand{\cM}{\mathcal{M}}

\newcommand{\cZ}{\mathcal{Z}}

\newcommand{\itPi}{\mathit{\Pi}}
\newcommand{\mg}{\mathcal{MG}}

\newcommand{\A}{\mathcal{A}}
\newcommand{\playerset}{\mathscr{N}}

\newcommand{\NashConv}{\textsc{SumRegret}\xspace}
\newcommand{\Regret}{\textsc{Regret}\xspace}

\newcommand{\ie}{{\it i.e.}}

\newcommand{\eg}{{\it e.g.}}
\newcommand{\lnorm}[1]{\lvert #1\rvert}

\newcommand{\Barg}{\texttt{Barg}} 
\newcommand{\perm}{\mathit{perm}}
\newcommand{\GumbelScore}{\mathit{GS}}
\newcommand{\Improve}{\mathit{Imp}}

\newcounter{lzNoteCounter}

\usepackage{newfloat}
\usepackage{listings}

\newtheorem{theorem}{Theorem}

\title{A Meta-Game Evaluation Framework for Deep Multiagent Reinforcement Learning}

\author{
Zun Li$^1$
\and
Michael P. Wellman$^2$
\affiliations
$^1$Google DeepMind,
$^2$University of Michigan\\
\emails
lizun@google.com,
wellman@umich.edu
}

\begin{document}

\maketitle

\begin{abstract}
Evaluating deep multiagent reinforcement learning (MARL) algorithms is complicated by stochasticity in training and sensitivity of agent performance to the behavior of other agents.
We propose a meta-game evaluation framework for deep MARL, by framing each MARL algorithm as a \emph{meta-strategy}, and repeatedly sampling normal-form empirical games over combinations of meta-strategies resulting from different random seeds.
Each empirical game captures both \emph{self-play} and \emph{cross-play} factors across seeds.
These empirical games provide the basis for constructing a sampling distribution, using bootstrapping, over a variety of game analysis statistics.
We use this approach to evaluate state-of-the-art deep MARL algorithms on a class of negotiation games.
From statistics on individual payoffs, social welfare, and empirical best-response graphs, we uncover strategic relationships among self-play, population-based, model-free, and model-based MARL methods.
We also investigate the effect of run-time search as a \emph{meta-strategy operator}, and find via meta-game analysis that the search version of a meta-strategy generally leads to improved performance.
\end{abstract}

\section{Introduction}


Evaluating complex AI algorithms requires careful attention to stochastic factors.
Deep reinforcement learning (RL) algorithms in particular are subject to randomness within the algorithm and the operational environment, and variations with choices of hyperparameters and initial conditions.
It is conventional to address these uncertainties in part by aggregating results across multiple runs \citep{bellemare2013arcade,machado2018revisiting}.
Deep multiagent RL (MARL) algorithms present these issues, and additional challenges due to agent interactions.
Evaluating performance against humans has been one source of compelling demonstrations \citep{silver2016mastering,Vinyals19Alphastar,wurman2022outracing,perolat2022mastering}, but this approach is limited by the range of tasks for which human expertise exists, and the cost of engaging it when it is available.
Generally speaking, we lack evaluation protocols for comparing different MARL methods in a statistically principled way.

In purely adversarial (\ie, two-player zero-sum) environments, distance to Nash equilibrium may be a sufficient metric \citep{brown2020combining,schmid2023student}, as all equilibria are interchangeably optimal.
More generally, where there are multiple equilibria or where we do not necessarily expect equilibrium behavior, the metrics for MARL performance may be less clear.
In collaborative domains, global team return is the common objective \citep{foerster2018counterfactual, rashid2020monotonic}, however complex learning dynamics may lead agents using the same MARL algorithm to equilibria of distinct machine conventions in different runs \citep{hu2020other}.

We seek an approach to evaluating deep MARL in general-sum domains.
We propose a \emph{meta-game} evaluation framework (\S\ref{sec:framework}), which frames MARL algorithms as \emph{meta-strategies}: mapping games and random seeds to joint policies.
We sample seed combinations to generate meta-game instances, from which we compute evaluation metrics of interest based on game-theoretic solution concepts. 
Through resampling and bootstrap techniques, we generate a statistical characterization of algorithm performance in these games. 
Our contributions:
\begin{itemize}
    \item The meta-game evaluation framework.
    \item A new search algorithm for imperfect information games, \textit{Gumbel Information-Set Monte-Carlo Tree Search} (\S\ref{alg:gumbel-search}), based on the recent development of Gumbel AlphaZero \citep{danihelka2022policy}.
    \item Extensive experiments on state-of-the-art MARL algorithms (\S\ref{sec:benchmarks}) on a class of negotiation games (\S\ref{sec:domain}), illustrating the framework and providing new evidence regarding the algorithms studied.
\end{itemize}

\section{Related Work}
While evaluating single-agent deep RL algorithms is well-studied \citep{henderson2018deep,jordan2020evaluating,agarwal2021deep}, there are relatively few works that consider evaluation principles for deep MARL \citep{gronauer2022multi}.
Typically, agents are evaluated against a selected set of background opponents or emergent behaviors in certain contexts \citep{lowe2017multi,leibo2021scalable}.
Other work has defined evaluation protocols for cooperative settings
\citep{papoudakis2020comparative,gorsane2022towards}, where a global team reward is well-defined. 
Our scenario falls into the \textit{agent-vs-agent} category of \citet{balduzzi2018re}, who argue for the necessity of comprehensively considering the possible joint interactions.

\citet{kiekintveld2008selecting} employed a concept of meta-games for evaluating general game-playing methods. 
In their setting, meta-strategies map normal-form game specifications to strategy choices.
\citet{treutlein2021new} construct \textit{label-free coordination} (LFC) games among instances of a cooperative MARL algorithm, in order to study zero-shot coordination \citep{hu2020other}.
LFC games are a kind of meta-game in our sense, as they take dynamic game descriptions (Dec-POMDPs) as input.

Bootstrapping is a non-parametric statistical approach that constructs sampling distributions for any specified statistic by resampling from the original dataset \citep{davison1997bootstrap}.
Bootstrapping techniques have been applied in machine learning methods such as aggregating decision-tree models \citep{breiman1996bagging}.
\citet{wiedenbeck2014bootstrap} applied the bootstrap for statistical analysis of game-theoretic models estimated from simulations.





\section{Game Theory Preliminaries}

A \term{normal-form representation} of a \term{game}~$\cG$ consists of a \term{player set} $\playerset=\{1, \dotsc, N\}$, and for each player $i\in\playerset$ a \term{pure strategy space} $\itPi_i$ and a \term{utility function} $u_i:\itPi\rightarrow \mathbb{R}$.
$\itPi=\itPi_1\times\dotsm\times\itPi_N$ is the \term{joint pure strategy space}.
A \term{mixed strategy} $\sigma_i\in\Delta(\itPi_i)$ for player~$i$ defines a probability distribution over that player's pure strategy space.
Player~$i$'s \term{payoff} for choosing $\sigma_i$ while others play $\sigma_{-i} = (\sigma_j)_{j\neq i}$ is given by expectation over the respective mixtures:
 $u_i(\sigma)=\mathbb{E}_{\pi_i\sim\sigma_i, \pi_{-i}\sim\sigma_{-i}}[u(\pi_i, \pi_{-i})]$.
 
The \term{regret} of player $i$ who plays $\sigma_i$ when others are playing $\sigma_{-i}$ is $\Regret_i(\sigma_i, \sigma_{-i})=\max_{\pi_i^\prime\in\itPi_i}u_i(\pi_i^\prime,\sigma_{-i})-u_i(\sigma_i,\sigma_{-i})$.
A \term{Nash equilibrium} (NE) is a mixed strategy profile $\sigma$ such that nobody has positive regret: $\forall i.\ \Regret_i(\sigma_i, \sigma_{-i})=0$.
As a measure of approximation to NE, we define $\NashConv(\sigma)=\sum_i\Regret_i(\sigma_i, \sigma_{-i})$.

If a game is known to be symmetric, we can reduce its normal-form description complexity.
Formally, in a \term{symmetric game}, players share the same strategy space $\itPi_1=\dotsm=\itPi_N=\itPi$, and utility function $u_1=\dotsm = u_N=u$.
Furthermore, each player's utility is permutation-invariant to other players' strategies.
For symmetric games we overload~$\itPi$ to refer to the common individual strategy space, rather than the joint space.
In a \term{symmetric profile}, every player adopts the same (generally mixed) strategy.
Solutions in symmetric profiles are guaranteed to exist in relevant settings \citep{nash1951non, cheng2004notes,hefti2017equilibria}, and are generally preferred absent any basis for breaking symmetry \citep{Kreps90a}.

Also given symmetry, let $u(\sigma^\prime, \bm{\sigma})$ be the expected payoff of a player when it chooses a strategy $\sigma^\prime\in\Delta(\itPi)$ while the other $N-1$  play the same mixed strategy $\sigma\in\Delta(\itPi)$.
We thus have $\Regret(\sigma^\prime, \sigma)=\max_{\pi^\prime\in\itPi}u(\pi^\prime, \bm{\sigma})-u(\sigma^\prime,  \bm{\sigma})$.
A symmetric NE strategy $\sigma$ satisfies $\Regret(\sigma, \sigma)=0$.

An \term{extensive-form game} (EFG) representation goes beyond normal form to include temporal and information structure.
Effectively, an EFG defines a dynamical system (usually visualized by a tree) with a world state or history $h$ that is not necessarily fully observable by every player.
At each state $h$, starting with the initial state $h_0$, there is a single player $\tau(h)\in\playerset\cup\{c\}$, where $c$ is the \term{chance player}, designated to select an action $a\in\A(h)$.
The chance player chooses according to a fixed random policy.
Player $i\in\playerset$ chooses according to its \term{information state/set}, $s_i(h)$, which comprises the information that $i$ has observed at~$h$.
Often $s_i(h)$ is represented by a concatenation of public and private action-observation histories.
Following action~$a$, the world transits to $h^\prime=ha$ until a terminal state $z\in\cZ$ is reached.
Then each player $i\in\playerset$ receives a utility $u_i(z)$ as a function of $z$.
A \term{behavioral strategy} or \term{policy}, $\pi_i$, of player~$i$ is a mapping from $i$'s infostates to distributions over actions.
For EFGs we overload $\itPi_i$ to refer to the set of behavioral strategies of player~$i$.
A joint behavioral strategy profile $\pi$ thus induces a probability distribution over the terminal states and we define $u_i(\pi)$ as the expected payoff for player~$i$ under this distribution.
In this paper, we consider EFGs with \term{perfect recall}, that is, information sets $s_i(h)$ distinguish all actions $i$ had taken to reach~$h$.
A consequence is that any mixed strategy $\sigma_i\in\Delta(\itPi_i)$ is payoff-equivalent to some behavioral strategy $\pi_i\in\itPi_i$ \citep{aumann1964mixed}.

\section{Multiagent Training Algorithms}

We define a \term{multiagent training algorithm} (MATA) $\cM$ as a stochastic procedure that produces a policy profile $\pi=\cM(\cG, \Theta, \omega)$ for an EFG.
In general, the input EFG $\cG$ cannot be tractably represented as an explicit game tree.
Instead, we assume the game is given in the form of a black-box simulator that the algorithm can exercise by submitting actions and receiving observations and rewards.
$\Theta$ is the set of hyperparameters of the algorithm, and
$\omega$ is a random seed.
If $\cG$ is symmetric, then it is often natural to constrain the output $\pi$ to be likewise symmetric (\ie, single policy to be played by all).
More generally, $\pi$ is a policy profile.
For the AlphaZero MATA, for example, $\cG$ could be represented by a Go simulator, and $\Theta$ would include the learning rate schedule and neural architecture.
The output profile $\pi$ specifies Go-playing policies for white and black, respectively.

A MATA is effectively a form of \term{meta-strategy}: a procedure that given a game $\cG$, generates a strategy profile for $\cG$.
We can employ the MATA to play from the perspective of any particular player~$i$, simply by selecting the $i^\text{th}$ element $\pi(i)$ of the output profile.

A key issue for analysis of MATAs is uncertainty in strategy generation.
It has been well observed (\eg, by \citet{hu2020other}) that a MATA with the same $\cG$ and $\Theta$ but different $\omega$ may generate policies with vastly different strategic behaviors.
For example, in a negotiation game, different runs of a MATA may lead to strategies that adopt distinct offering conventions.
In the present work, we assume the hyperparameters $\Theta^m$ for each MATA $\cM^m$ have been fixed, so the uncertainty in behavior of a training algorithm is fully captured by the random seeds.
Note that there is always discretion about what one considers a distinct MATA~$\cM$ versus a parametric variation, so it is possible to bring choice among hyperparameter settings within the scope of our analysis framework.

\section{Meta-Game Evaluation Framework}\label{sec:framework}

Given $M$ different MATAs $\{(\cM^1,\Theta^1),\dotsc,(\cM^M,\Theta^M)\}$ with associated hyperperameters, how can we evaluate their relative performance with respect to a given game $\cG$?
Viewing the MATAs simply as game-solvers, we could focus on measures of their effectiveness in deriving a solution---for example, time and accuracy of convergence to Nash equilibrium.
Viewing the MATAs' role as generating strategies to play a game, however, requires a different focus that considers the interaction with other strategy generators.
This is particularly salient for games that have a multiplicity of solutions (the general case), or for which the operable solution concept may be open to question.
Consequently, we propose to analyze competing MATAs by framing their interaction as itself a game. 
As MATAs are \textit{meta-strategies}, we refer to this approach as a \term{meta-game} evaluation framework.

As noted above, we are particularly concerned with uncertainty in the results of multiagent training.
In analysis of the MATA meta-game, therefore, we aim to characterize the implications of this uncertainty in probabilistic form.

\subsection{Empirical Game-Theoretic Analysis}
\label{sec:egta}

Our approach employs \term{empirical game-theoretic analysis} (EGTA), a methodology for reasoning about games through agent-based simulation \citep{Tuyls20,Wellman16}.
EGTA aligns with our assumption that the game of interest $\cG$ is defined by a black-box simulator.
In the typical framing, EGTA operates by estimating an \term{empirical game model} in normal form over an enumerated strategy set. 
The enumerated strategies are a small selection from the full strategy space of the extensive game represented by the simulator.

We employ EGTA with respect to an $N$-player game of interest~$\cG$, and strategies defined by the output of MATAs.
The meta-game is likewise over $N$ players, and is symmetric regardless of whether $\cG$ is symmetric or not.
Let $\hat{\pi}^m$ denote the output from MATA~$m$, for instance $\hat{\pi}^m=\cM^m(\cG,\Theta^m,\omega)$: the result from running the MATA on~$\cG$ for a particular random seed.
We also allow that $\hat{\pi}^m$ be an aggregate of policy profiles from multiple random seeds.
From these MATA-generated profiles $\hat{\itPi}=\{\hat{\pi}^1,\dotsc,\hat{\pi}^M\}$, we construct an empirical meta-game $\mg(\hat{\itPi})$ over policy space $\hat{\itPi}$ as follows.

If the base game $\cG$ is symmetric, then the $\hat{\pi}^m$ are single-player policies, and we estimate the meta-game utility function by the standard EGTA approach of simulating profiles over $\hat{\itPi}$.
If $\cG$ is not symmetric, then each $\hat{\pi}\in\hat{\itPi}$ is an $N$-player profile for $\cG$. 
Simulating a meta-game profile $(\hat{\pi}_1,\dotsc,\hat{\pi}_N)$ of these base-game profiles entails first assigning the $\hat{\pi}$ to players, according to a random permutation $\perm$ drawn uniformly from the $N!$ possibilities. 
Then it simulates a play where player~$i$ of $\cM\cG$ plays $\hat{\pi}_{i}(\perm(i))$ as if it is player $\perm(i)$ in $\cG$.
This construction also corresponds to an EFG beginning with a root chance node that uniformly chooses among $N!$ different outcomes, each followed by a copy of the original EFG with player indices permuted.


Symmetry of the meta-game reflects a view that, for multiagent training, developing effective strategies for each of the player positions is equally important.
If this were not the case, or if one wished to perform an analysis of the differential effectiveness of various MATAs from the perspectives of different players or roles, a non-symmetric (or role-symmetric) meta-game model could be constructed instead.

\subsection{Meta-Game Evaluation Procedure}
\label{sec:meta-game}

Just as single-agent RL algorithms are statistically evaluated by return performance across different random seeds, we can analyze strategic properties among MATAs across different \textit{combinations} of seeds.
Let $\bm{X}$ denote statistics characterizing the strategic properties of interest (discussed below).
Given an $N$-player game $\cG$ and parametrized MATAs $\{(\cM^1,\Theta^1),\dotsc,(\cM^M,\Theta^M)\}$, our evaluation procedure comprises of the following steps:
\begin{enumerate}
    \item Select a finite set of seeds $\Omega^m$ for each MATA~$m$.
    Generate $\hat{\pi}^m(\omega) = \cM^m(\cG,\Theta^m,\omega)$ for each $\omega\in\Omega^m$.
    \item For each $m$, uniformly sample $\lnorm{\Omega^m}$ seeds from $\Omega^m$ with replacement, yielding the sequence $(\omega_1,\dotsc,\omega_{\lnorm{\Omega^m}})$.
    Let $\hat{\pi}^m$ be a profile that 
    that is payoff-equivalent to a uniform mixture over the multiset $\{\hat{\pi}^{m}(\omega_1),\dotsc, \hat{\pi}^{m}(\omega_{\lnorm{\Omega^m}})\}$.
    \item Given $\hat{\itPi} = \{\hat{\pi}^1,\dotsc,\hat{\pi}^M\}$, estimate the symmetric empirical meta-game $\mg(\hat{\itPi})$ as described in \S\ref{sec:egta}.
    \item Compute the statistics-of-interest $\bm{X}$ from $\mg(\hat{\itPi})$
    \item Repeat Steps 2 through~4.
    Estimated profile payoffs should be memoized for reuse across iterations. 
    Obtain an empirical distribution of~$\bm{X}$ and report statistical properties of~$\bm{X}$.
\end{enumerate}


One way to understand this evaluation procedure is to view each MATA as a $\emph{mixed strategy}$, selecting policies uniformly over the possible random seeds.
The ``ground-truth" meta-game represents an expectation over the results of this randomization. 
The empirical meta-game estimates this from finite samples $\Omega^m$.
By resampling from these seeds at hand, Step 2 constructs multiple empirical games among MATAs, from which we construct sampling distributions over $\bm{X}$ using bootstrapping.

There are a variety of choices for statistics $\bm{X}$ to gather.
The only requirement is that $\bm{X}$ can be computed from the information in a normal-form game model.
For example, one possible metric on MATA performance is \term{uniform-score}: average payoff against a uniform distribution over other MATAs.
Such scores have been employed in a variety of contexts, however putting equal weight on the possible counterparts is questionable, as they are not equally relevant \citep{balduzzi2018re}.

An alternative proposed by \citet{jordan2007empirical} is \term{NE-regret}: $\Regret(\pi, \sigma^*)$,
where $\sigma^*$ is a symmetric mixed equilibrium of $\mg(\hat{\itPi})$.
The motivation for this measure is that it focuses on behavior against rational opponents.
Performance against obviously flawed opponents should carry much less weight, as they are less likely to be encountered in practice, all else equal.%
\footnote{If however there is some basis to expect opponents who are flawed or boundedly rational in some particular way, then by all means it would make sense to measure regret with respect to a solution concept capturing that basis.}
For games with multiple equilibria, however, NE-regret is sensitive to the choice of solutions $\sigma^*$.
This sensitivity is inherent to situations with multiple equilibria, but it can still be helpful to adopt a focal equilibrium to reduce ambiguity in analysis.
\citet{balduzzi2018re} proposed \term{Nash averaging}, which is essentially NE-regret with respect to the max-entropy equilibrium.
The intuition for preferring to evaluate with respect to higher-entropy solutions is that they reflect diversity, and thus reward robustness to a wide range of rational opponents.


\subsection{Max-Entropy Nash Equilibrium}

Computing max-entropy NE is hard in general, but practically feasible to approximate for bi-matrix games of modest size.
We adapt the mixed-integer programming formulation of \citet{sandholm2005mixed}:
\begin{equation}
\begin{aligned}
\min_{\sigma^*} \quad & \sum_{\pi\in \hat{\itPi}}\sigma^*(\pi)\log\sigma^*(\pi)\\
\textrm{s.t.} \quad
 &\forall \pi\in \hat{\itPi}. \quad
    u_{\pi}= \sum_{\pi^\prime\in\hat{\itPi}}\sigma^*(\pi^\prime)u(\pi, \bm{\pi}^\prime),\\
     &u^*\ge u_{\pi},
    \quad u^*-u_{\pi}\le Ub_{\pi},\quad
    \sigma^*(\pi)\le 1-b_{\pi},\\
  &\sigma^*(\pi)\geq0, \quad
  \sum_{\pi}\sigma^*(\pi)=1,
  \quad b_{\pi}\in\{0, 1\},\\
\end{aligned}
\label{prog: mip}
\end{equation}
with real variables $\sigma^*(\pi)$, $u_{\pi}$, binary variables $b_{\pi}$ for each strategy~$\pi$, and an equilibrium payoff real variable~$u^*$.
$U$~is the maximum difference across payoffs.
The variables $b_{\pi}$ indicate whether strategy~$\pi$ is outside the equilibrium support.
That is, $b_{\pi}=1$ iff $\sigma^*(\pi)=0$.
Otherwise $u_{\pi}=u^*$: strategies in the support have the same payoff.
The details of our implementation are in App.~\ref{appendix:max-ent}.

Although max-entropy NE is not generally unique, it narrows the possibilities considerably compared to unconstrained (approximate or exact) NE.



\section{Search as a Meta-Strategy Operator}
\label{alg:gumbel-search}
Many MATAs produce a policy network $\bm{p}$ that maps directly from an infostate to a distribution over actions in a forward pass for every player.
Recent work has found that leveraging computation at run-time and adding search to $\bm{p}$ can improve performance in large EFG domains \citep{silver2018general,brown2020combining,schmid2023student}.
As a case study for our meta-game evaluation framework, we apply it to investigate the effect of search as a general policy improver.

Toward that end, we propose a heuristic search method for large EFGs based on information-set MCTS (IS-MCTS) \citep{CowlingISMCTS} and Gumbel AlphaZero \citep{danihelka2022policy}.
Alg.~\ref{algorithm: g-ismts} presents the procedure, \term{Gumbel IS-MCTS}, in detail.
Parameterized by a policy net $\bm{p}$ and a value net $\bm{v}$, Gumbel IS-MCTS conducts multiple passes over the game-tree guided by $\bm{v}$ and $\bm{p}$ at an input infostate~$s$, and outputs an action $a$ for decision-making.
We can apply this procedure to a variety of underlying MATAs, as a \term{meta-strategy operator}: transforming $\cM$ to $\cM'$ (with additional hyperparameters like simulation budget).
The meta-strategy~$\cM'$ in effect adds run-time search to the output policy $\bm{p}$ of~$\cM$.
Unlike AlphaZero---which uses the same MCTS method for training and run-time, with meta-strategy operators we can explore a variety of MATAs as training-time methods which produce $\bm{v}$ and $\bm{p}$ for search at test-time \citep{sokota2023update} (details in~\S\ref{subsec: all-search}).
\begin{algorithm}[tb]
   \caption{Gumbel IS-MCTS}
   \label{alg:is-mcts-br}
\begin{algorithmic}[1]
\FUNCTION{\texttt{Gumbel-Search}($s$, $\bm{v}, \bm{p}$)}
   \STATE $\forall (s, a).\ R(s, a)\gets 0, C(s, a)\gets 0$\;
   \STATE $\forall a\in\cA(s)$. sample $g(a)\sim \mathit{Gumbel}(0), \hat{\cA}\leftarrow\cA(s)$ 
   \REPEAT
   \STATE Sample a world state: $h \sim \Pr(h\mid s, \bm{p})$ \label{line:g-sample}
   \WHILE{}
   \IF{$h$ is terminal}
   \STATE $\bm{r}\gets$ payoffs of players {\bf Break}
   \ELSIF{$i\triangleq\tau(h)$ is chance}
   \STATE $a \gets$ sample according to chance
   \ELSIF{$s_{i}(h)$ not in search tree}
   \STATE Add $s_{i}(h)$ to search tree
   \STATE $\bm{r}\leftarrow \bm{v}(s_{i}(h))$. {\bf Break}
   \ELSIF{$s_{i}(h)$ is root node $s$}
   \STATE $a, \hat{\cA} \gets $ one step of sequential halving (Alg.~\ref{algorithm: sh}) based on  $\GumbelScore(s, a)$ and remaining actions in $\hat{\cA}$
   \ELSE
   \STATE Select $a$ according to Eq.(\ref{eq: non-root}) in App.~\ref{sec: non-root}.
   \ENDIF
   \STATE $h\leftarrow ha$
   \ENDWHILE
   \FOR{$(s_{i}, a)$ in this trajectory}
   \STATE Increment $R(s_i, a)$ by $\bm{r}_i$,  $C(s_i, a)$ by 1.
   \ENDFOR
   \UNTIL{$num\_sim$ simulations done}
   \STATE {\bf return} Action $a$ that remains in $\hat{\cA}$
 \ENDFUNCTION
 \label{algorithm: g-ismts}
\end{algorithmic}
\end{algorithm}

Just like MCTS, IS-MCTS incrementally builds and traverses a search tree and aggregates statistics such as visit counts $C(s, a)$ and aggregated values $R(s, a)$ for visited $(s, a)$ pairs.
During each simulation of the search (line~5), a world state is sampled from a posterior belief $\Pr(h\mid s, \bm{p})$ assuming the opponents played according to $\bm{p}$ prior to $s$.
In our test domains, $\Pr(h\mid s, \bm{p})$ can be computed exactly via Bayes's rule, where in larger domains, using particle filtering \citep{silver2010monte} or deep generative models \citep{hu2021learned,li2023combining} to approximate $\Pr(h\mid s, \bm{p})$ are possible.
Further technical details are provided in App.~\ref{sec: detail-g-search}.

The key feature of Gumbel IS-MCTS is how it selects actions at the search nodes.
At the beginning of the search (line~3), a Gumbel random variable, $g(a)$, is sampled i.i.d. for each legal action $a$ of the root, for later use in action selection.
At the root (line~15), the algorithm treats each legal action as an arm of a stochastic bandit, and uses a sequential-halving algorithm \citep{pepels2014minimizing} (Alg.~\ref{algorithm: sh}) to distribute the simulation budget.
Sequential-halving algorithms usually are designed for minimizing the \emph{simple regret} \citep{bubeck2009pure}, which is the regret at the last-iteration action recommendation.
By contrast, UCB-style algorithms are usually designed for minimizing the \emph{accumulated regret} during an online learning process.
For a game-playing search algorithm, minimizing simple regret makes more sense in terms of producing a single optimal action at a decision point.

We assign to each arm $a$ a \term{Gumbel score} $\GumbelScore(s, a)=
g(a)+\text{logit } \bm{p}(s, a)+G(\hat{q}(s, a))$.
The second term is the logit of $a$ produced by $\bm{p}$, and the third term is a monotone transformation of the action value $\hat{q}(s, a)$, which is estimated by $R(s, a)$, $C(s, a)$, and $\bm{v}$.
The intuition is that a high $\hat{q}(s, a)$ value indicates a direction for policy improvement.
Indeed, the improved policy $\Improve(\bm{p})(s, a)\triangleq\text{SoftMax}(\text{logit } \bm{p}(s, a)+G(\hat{q}(s, a)))$ provably achieves higher expected values \citet[App.~B]{danihelka2022policy}.
The forms of $G$ and $\hat{q}$ is detailed in App.~\ref{sec: qhat}.

Adding Gumbel noise $g(a)$ implements the ``Gumbel top-K-trick'': deterministically selecting the top $K$ actions according to $\GumbelScore(s, a)$ is equivalent to sampling $K$ actions from $\Improve(\bm{p})(s, a)$ without replacement \citep{huijben2022review}.
The Gumbel score induces a low-variance non-deterministic action selection of the root node during the sequential halving process, which encourages exploration while distributing the simulation budget toward actions likely to yield higher expected values.

At a non-root node (line 17), an action is selected to minimize the discrepancy between 
$\Improve(\bm{p})$ and the produced visited frequency (details in App.~\ref{sec: non-root}).
At the end of the search, Gumbel IS-MCTS outputs the action~$a$ that survives the sequential halving procedure.



\section{Evaluation Study}

\subsection{Domain: Alternating Negotiation}\label{sec:domain}


\begin{figure}[t]
    \centering
    \includegraphics[width=0.9\columnwidth]{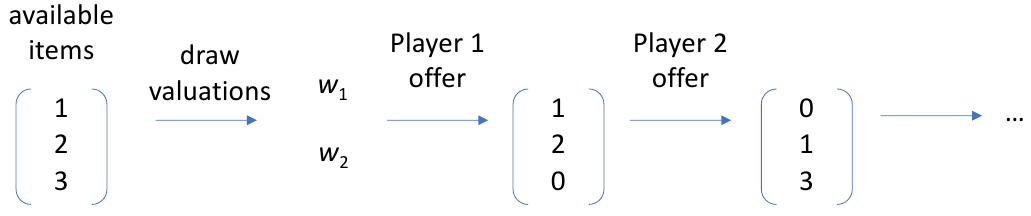}
\caption{Example start of sequential bargaining game instance.}
    \label{fig:dond-example}
\end{figure}

We test our MATAs and evaluation framework in the context of a two-player negotiation game. 
The particular version we study, called ``Deal-or-No-Deal'' (DoND) by \citet{Lewis17}, is a family of two-player negotiation games where the players alternate proposals to divide a pool of resources: three different goods available in various amounts.
At the start of the game, quantities of goods in the pool and players' private valuations for units of each good are sampled from a known distribution.
The valuation distribution is constrained to fix a constant value for both players of the aggregate pool: $\mathbf{c}\cdot\mathbf{w}_1 = \mathbf{c}\cdot\mathbf{w}_2 = 10$.
For example, in Fig.~\ref{fig:dond-example}, the pool of goods can be represented as $\mathbf{c}=[1,2,3]$.
The negotiation proceeds through alternating proposals for dividing the resources.
In this example, player 1 first proposes $\bm{o}_1=[1, 2, 0]$ for itself which entails $\bm{o}_2=\mathbf{c}-\bm{o}_1=[0,0,3]$ for player~2.
Player~2 rejects and proposes a counter-offer, to which 1 may either agree or continue bargaining.
For a particular game parametrization $\Barg(T, \varepsilon, \gamma)$, the game ends if either (1) a deal is made, (2) a maximum number of rounds $T$ is reached, or (3) chance decides to terminate the game, which happens at every round with probability~$\varepsilon$.
If an agreement $(\bm{o}_1,\bm{o}_2)$ was reached at the $t^\text{th}$ round, player~$i$ receives payoff $\gamma^t\bm{w}_i\cdot\bm{o}_i$.
Otherwise both players receive zero payoff.

DoND is a family of challenging general-sum environments with imperfect information. 
Players have a common interest in reaching a deal (and quickly, for $\gamma<1$), and the different ways of dividing the pool have different total value.
In our studies, we sample configurations uniformly from a database of 6796 published by \citet{Lewis17}.
This leads to prohibitive complexity: a game with $T=10$ has $O(10^{11})$ infosets for every player, which is intractable for enumeration (detailed in App.~\ref{sec:app-games-dond}), and grows exponentially with~$T$.

\subsection{Benchmark Algorithms}\label{sec:benchmarks}

Our meta-game evaluation considers the following $M=17$ MATAs.
These algorithms represent a comprehensive set of state-of-the-art MARL algorithms, including methods using self-play based training, population based training, and model-free and model-based approaches.
Details are below and in App.~\ref{sec:hyper}.

\subsubsection{Independent/Multiagent PPO (IDPPO/MAPPO)}
Both IDPPO and MAPPO train policy and value nets using self-play trajectories by 
minimizing the trust-region clipped loss \citep{schulman2017proximal} using the generalized advantage estimator (GAE) \citep{mnih2016asynchronous}.
Value nets are trained by minimizing L2 loss from the targets produced by GAEs.
In IDPPO each player maintains its own policy and value nets whereas in MAPPO \citep{yu2022surprising} players share the same neural nets.


\subsubsection{Regularized Nash Dynamics (R-NaD) and NFSP}
Both R-NaD \citep{perolat2021poincare} and Neural Fictitious Self-Play (NFSP) \citep{heinrich2016deep} are self-play model-free MARL algorithms originally designed for purely adversarial settings.
R-NaD has recently shown success in producing human-level agents in Stratego \citep{perolat2022mastering},
where it iteratively trains policy nets by minimizing NeuRD loss \citep{hennes2020neural} and value nets using the V-Trace estimator \citep{espeholt2018impala} on a sequence of regularized games. 
NFSP mimics the classic fictitious play algorithm \citep{heinrich2016deep} by alternating between training a supervised learning net that summarizes historical plays and training a DQN policy that serves as a best response.

\subsubsection{Policy Space Response Oracles (PSRO) and FCP}
PSRO \citep{lanctot2017unified} and Fictitious Co-Play (FCP) \citep{strouse2021collaborating} are two population-based MARL algorithms.
PSRO is an EGTA method that iteratively adds policies that are best responses to the current solution. 
We use max-entropy Nash as the solution concept (solving an asymmetric version of~\eqref{prog: mip}), and PPO as the best-response method.
We consider two ways of extracting the final agents: (i)~PSRO: using the final-iteration solution and (ii)~PSRO-LAST: using the final-iteration best-response policies.
FCP has notably demonstrated success in collaborating with humans from scratch \citep{strouse2021collaborating}.
FCP builds a population by picking policies of different skill levels on multiple self-play runs, and trains final best responses against such populations.
Our implementation of FCP uses IDPPO to generate self-play runs, picks the agents based on social welfare, and uses PPO as the best-response method.


\subsubsection{Gumbel Search and Vanilla AlphaZero-style Search}\label{subsec: all-search}
We include the following variants of Gumbel IS-MCTS:
(i)~G-Search: Using Gumbel IS-MCTS for both training the networks (Alg.~\ref{algorithm: self-play}) at training time and conducting search at test time.
(ii)~G-Search-PN: Using search at training time, but only policy net without search at test time.
(iii)~G-Search-IDPPO: Using IDPPO to train the policy and value nets, based on which Gumbel IS-MCTS executes test-time search. 
(iv)~G-Search-MAPPO and (v)~G-Search-R-NaD are similarly defined.
In addition, we implement an extension of the original AlphaZero-style MCTS to IS-MCTS.
The differences between this search method (Alg.~\ref{algorithm: va-mcts} in App.~\ref{sec:pseudocode}) and Gumbel IS-MCTS are in the mechanisms for exploration (\eg, Dirichlet noise) and action selection (\eg, PUCT) within the search tree.
We include (vi)~VA-Search: Using this search for both training time and test time, and (vii)~VA-Search-PN: similarly defined as (ii).

\subsubsection{Heuristic Strategies}
We further include the following baseline bargaining strategies: (i)~{\bf Uniform}, which uniformly samples among all legal actions at each decision point. (ii)~{\bf Tough}, which never agrees, and always proposes uniformly among offers that maximize its own payoff. (iii)~{\bf Soft}, which always agrees, or uniformly proposes among all offers if it is the first-mover.

\begin{table*}[t!]
\begin{center}
\begin{adjustbox}{max width=\textwidth}
\begin{tabular}{|c|c|c|c||c|c|c|c|}
\hline
MATA Name & NE-Regret-Score & Uniform-Score & NE-NBS & MATA Name & NE-Regret-Score & Uniform-Score & NE-NBS\\
\hline
G-Search-R-NaD & {\bf 0.002$\pm$0.020} & 6.158$\pm$0.040 & 44.010$\pm$0.794 & MAPPO  & 0.721$\pm$0.059 & 5.602$\pm$0.035 & 32.832$\pm$0.818\\
\hline
G-Search &  0.010$\pm$0.046 & {\bf 6.215$\pm$0.015} & {\bf 44.345$\pm$0.496} & IDPPO &  0.766$\pm$0.133 & 5.645$\pm$0.050 & 30.240$\pm$2.051\\
\hline
R-NaD &  0.045$\pm$0.043 & 5.977$\pm$0.027 & 42.008$\pm$0.738 & NFSP &  0.806$\pm$0.050 & 5.766$\pm$0.014 & 36.534$\pm$0.721\\
\hline
VA-Search &  0.092$\pm$0.023 & 6.074$\pm$0.029 & 40.836$\pm$0.551 & VA-Search-PN &  0.843$\pm$0.071 & 5.570$\pm$0.026 & 36.540$\pm$0.717\\
\hline
G-Search-MAPPO  & 0.354$\pm$0.047 & 6.038$\pm$0.027 & 37.808$\pm$0.962 & FCP  & 1.116$\pm$0.096 & 5.804$\pm$0.013 & 38.665$\pm$1.092\\ 
\hline
PSRO-LAST  & 0.414$\pm$0.051 & 5.890$\pm$0.026 & 39.187$\pm$0.762 & Tough  & 1.801$\pm$0.093 & 4.017$\pm$0.031 & 6.581$\pm$0.137\\
\hline
PSRO &  0.417$\pm$0.055 & 5.904$\pm$0.022 & 38.288$\pm$0.873 & Soft  & 3.189$\pm$0.098 & 3.773$\pm$0.013 & 27.023$\pm$0.855\\
\hline
G-Search-IDPPO  & 0.480$\pm$0.062 & 6.088$\pm$0.031 & 36.385$\pm$1.317 & Uniform  & 4.274$\pm$0.062 & 2.391$\pm$0.004 & 11.980$\pm$0.494\\
\hline
G-Search-PN  & 0.575$\pm$0.048 & 5.811$\pm$0.013 & 42.174$\pm$0.493 & & & &\\
\hline
\end{tabular}
\end{adjustbox}
\end{center}
\caption{Results for $\Barg(10, 0, 1)$, with 95\% confidence intervals.
MATAs are listed in increasing order of NE-Regret.
\label{tab:barg-10}}
\end{table*}

\subsection{Experimental Setup}
All methods are implemented within the programming model of OpenSpiel \citep{lanctot2019openspiel}.
Each MATA that involves neural training uses models of approximately equal size (\eg, number of layers, hidden nodes).
The neural net inputs are likewise the same across all MATAs, and include complete history information (\ie, losslessly represent infosets) for the DoND game.
For procedures that involve self-play or RL training we employ roughly the same number of training trajectories.
For all search methods we use 200 simulations per call.
We fine-tuned the learning rates based on $\NashConv$ performance.
Most of the MATAs reliably produce approximate equilibria. Details are in App.~\ref{sec:nashconv}.

We set $\lnorm{\Omega^m}=10$, $m\in \{1,\dotsc,M\}$.
To simplify the evaluation procedure, we pre-compute an empirical payoff matrix covering every policy pair: $(\hat{\pi}^{m}_1(\omega),\hat{\pi}^{m'}_2(\omega'))$, for $m,m'\in \{1,\dotsc,M\}$, $\omega$ and $\omega'$ among the seeds sampled for $\cM^{m}$ and $\cM^{m'}$, respectively.
For each payoff entry we run $2\times10^4$ simulations to compute the expected payoff.
We then sample and analyze $10^6$ $M\times M$ empirical meta-games from this matrix, per Steps 2--4 of \S\ref{sec:meta-game}, to obtain distributions over the statistics of interest.

\begin{table*}[t!]
\begin{center}
\begin{adjustbox}{max width=\textwidth}
\begin{tabular}{|c|c|c|c||c|c|c|c|}
\hline
MATA Name & NE-Regret-Score & Uniform-Score & NE-NBS & MATA Name & NE-Regret-Score & Uniform-Score & NE-NBS\\
\hline
G-Search-R-NaD &  {\bf 0.000$\pm$0.001} & {\bf 6.147$\pm$0.013} & 43.681$\pm$0.497 & PSRO &  0.488$\pm$0.056 & 5.688$\pm$0.052 & 39.851$\pm$0.728\\
\hline
G-Search &  0.005$\pm$0.033 & 6.078$\pm$0.019 & 43.198$\pm$0.545 & G-Search-MAPPO &  0.729$\pm$0.095 & 5.656$\pm$0.056 & 32.983$\pm$0.994\\
\hline
R-NaD &  0.045$\pm$0.010 & 6.121$\pm$0.012 & {\bf 44.072$\pm$0.417} & IDPPO &  0.749$\pm$0.059 & 5.504$\pm$0.042 & 36.350$\pm$0.685\\
\hline
VA-Search &  0.075$\pm$0.023 & 6.058$\pm$0.014 & 42.605$\pm$0.610 & PSRO-LAST  & 0.906$\pm$0.213 & 5.391$\pm$0.161 & 35.723$\pm$2.119\\
\hline
G-Search-PN & 0.294$\pm$0.019 & 5.833$\pm$0.016 & 42.285$\pm$0.448 & MAPPO &  1.135$\pm$0.144 & 5.306$\pm$0.086 & 31.878$\pm$1.378\\
\hline
FCP  & 0.360$\pm$0.026 & 5.864$\pm$0.013 & 43.321$\pm$0.265 & Soft  & 1.918$\pm$0.018 & 4.352$\pm$0.011 & 35.990$\pm$0.267\\
\hline
VA-Search-PN  & 0.412$\pm$0.033 & 5.668$\pm$0.014 & 40.314$\pm$0.325 & Uniform  & 4.161$\pm$0.031 & 2.370$\pm$0.007 & 12.562$\pm$0.211\\
\hline
NFSP &  0.421$\pm$0.011 & 5.786$\pm$0.014 & 39.151$\pm$0.481 & Tough  & 4.542$\pm$0.082 & 2.685$\pm$0.025  & 2.270$\pm$0.097 \\
\hline
G-Search-IDPPO & 0.450$\pm$0.038 & 5.820$\pm$0.019 & 36.997$\pm$0.432 & & & &  \\
\hline
\end{tabular}
\end{adjustbox}
\end{center}
\caption{Results for $\Barg(30, 0.125, 0.935)$, with 95\% confidence intervals.
MATAs are listed in increasing order of NE-Regret.
\label{tab:barg-30}}
\end{table*}

\subsection{Results}

We test all 17 MATAs on two DoND instances: \Barg(10, 0, 1) and \Barg(30, 0.125, 0.935).
The two settings are qualitatively different.
The latter includes discounting and per-round ending probability, incentivizing the players to find and agree on a good deal in early rounds. 
Without these factors, the first game reduces to an ultimatum-game-like environment.
Meta-game statistics are reported in Tables \ref{tab:barg-10} and~\ref{tab:barg-30}.

In addition to NE-Regret and uniform scores, we also report a statistic we term the \term{NE-Nash-Bargaining-Score} (NE-NBS).
The NE-NBS for MATA~$m$ in an empirical game is defined as the utility-product between $\pi^m$ and an opponent strategy $\sigma^*$: $u(\pi^m, \bm{\sigma}^*)\cdot u(\sigma^*, \bm{\pi^m})$.
Here $\sigma^*$ is a max-entropy Nash computed by~\eqref{prog: mip}.
This statistic is intended to measure the effectiveness of an agent in achieving high social welfare and fairness with a rational opponent.

From the tables we can see that while there is positive correlation between NE-Regret Scores and Uniform Scores, NE-Regret Scores are better at identifying the most robust MATAs:
the rankings of the top four MATAs in NE-regret scores are the same for both environments, and from the empirical distribution plots in App.~\ref{sec:regret-plots} we can see G-Search-R-NaD and G-Search consistently produce strategies that are selected in a max-entropy equilibrium support with high probability.
By contrast uniform-score comparisons can be distorted by non-salient success against weak strategies.
It is also interesting to note positive correlation between NE-Regret and NE-NBS.
This reflects the non-zero-sum nature of these environments.

The performance of search-based agents is especially noteworthy.
Three of the top four MATAs employ search at test time.
All search methods are stronger than their policy net counterparts, confirming that IS-MCTS is a policy-improver in imperfect information games.
We hypothesize that this is because a search-based algorithm is usually a better responder to its policy network counterpart, while both behave strategically similarly against other methods.
The relative strength among search-based algorithms mirrors that of their policy-net counterparts.
In both environments, G-Search and VA-Search are among the top four.
This suggests that using search at training time is compatible with the same search at test time, and produces better policy and value nets compared with other methods.

We notice some algorithms behave qualitatively differently in these two environments.
We found FCP consistently produces collaborative strategies with rather high agreement probability.
This may explain its better performance in the second environment, which encourages settling a deal in early rounds.
Likewise, Soft achieves higher individual scores and NE-NBS in the second environment.
By contrast, while Tough achieves better individual performance than Soft in the ultimatum game, it is still worse than all learning methods, as shown by its terrible NE-NBS values for both environments.



\begin{figure}[ht!]
    \centering
\includegraphics[height=3.55cm]{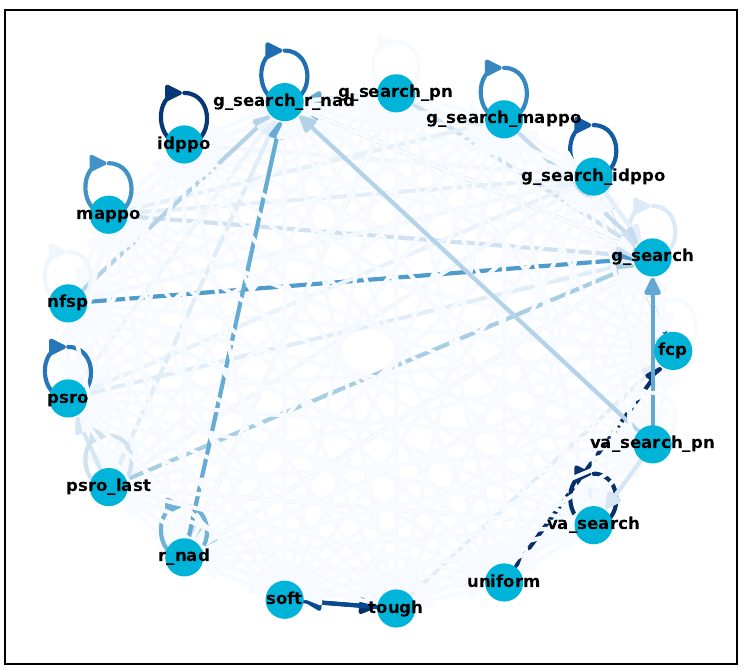}
\includegraphics[height=3.55cm]{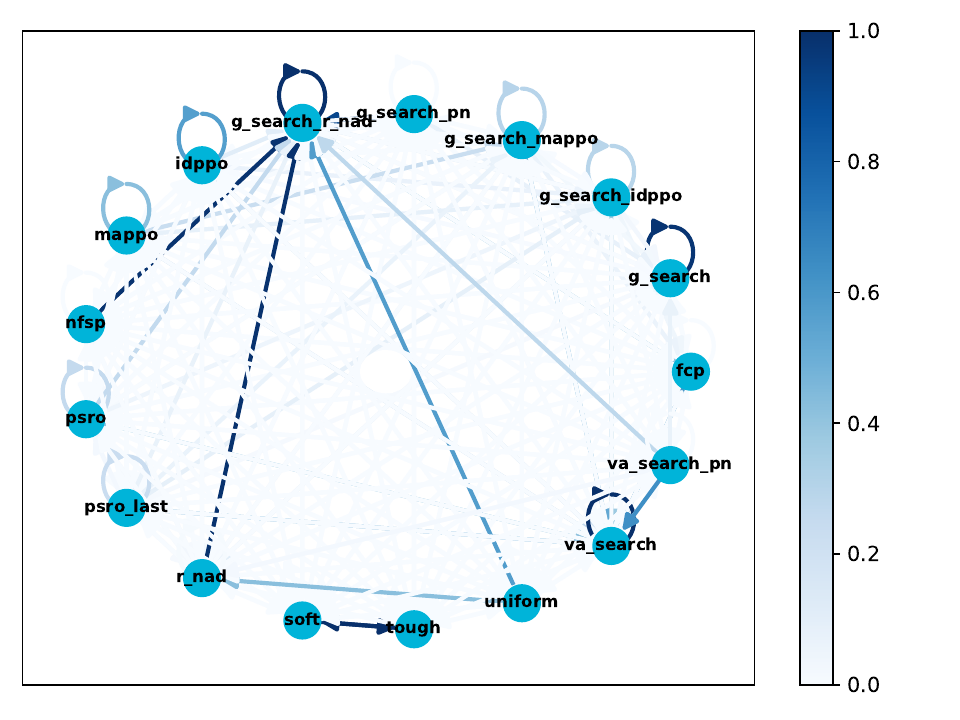}
    \caption{Empirical best-response graphs for \Barg(10, 0, 1) and \Barg(30, 0.125, 0.935).}
    \label{fig:br-graphs}
\end{figure}

Another interesting view is provided by \term{best-response graphs}.
In a best-response graph for a game instance, strategies are vertices, and there is a directed edge $(m_1\rightarrow m_2)$ iff $m_2=\argmax_{m^\prime} u(\pi^{m^\prime}, \bm{\pi^{m_1}})$.
We generate an aggregate graph for a MATA meta-game by recording the frequency of such edges in the sampled empirical games.
These are shown for the two game settings in Fig.~\ref{fig:br-graphs}.

The empirical best-response graphs support several observations. 
First, many MATA vertices such as IDPPO and MAPPO have self-edges with non-negligible frequencies.
This is consistent with previous observations that self-play MARL algorithms are likely to produce agents that overfit to themselves \citep{hu2020other}.
However, self-edges do not necessarily suggest poor generalization performances.
For example, G-Search-NaD has a self-edge with a high probability in both environments, but still performs the best.
Self-edges persist even after applying search as policy-improver at test time.
For example, there are self-edges for G-Search-IDPPO and G-Search-MAPPO.
This may suggest that there are certain strategic correlations between a search agent and its policy net counterpart.
Interestingly, we also observe certain self-edge probabilities for PSRO and PSRO-LAST, which are normally regarded as population-based training methods.
It could be that the equilibria selected at each iteration introduce correlations between different players.
By contrast, NFSP and FCP are two MATAs without strong self-edges; both  use a uniform distribution for opponent modeling, which may reduce the correlation between player positions.

Edges across MATAs illuminate strategic interaction structure.
Tough is unsurprisingly the best response to Soft.
G-Search-NaD is a strong attractor in both graphs, as G-Search is in the first.
Sometimes a search-based MATA is a best response to its policy-net counterpart, as illustrated by R-NaD to G-Search-R-NaD, and VA-Search-PN to VA-Search.
This could also explain why MATAs such as G-Search-PN and VA-Search-PN do not have apparent self-edges.



\section{Conclusion}

We propose a meta-game evaluation framework for MARL in general-sum environments. 
Our approach is analogous to the evaluation process for single-agent RL, effectively aggregating the strategic analysis procedures across possible worlds defined by different seed combinations.
We illustrated the method by constructing a meta-game model over a comprehensive set of multiagent training algorithms using deep RL on a class of negotiation games.
The meta-game analysis evaluated the algorithms in multiple ways, most prominently through max entropy NE-regret ranking and the structure of best-response graphs.
Bootstrap statistics provide a basis for assessing uncertainty in evaluation results.

Experimental results support several interesting observations about Gumbel IS-MCTS as a meta-strategy operator,  especially regarding the value of search at both training time and test time, and on patterns of strategic interactions among the algorithms.

A key feature of our evaluation framework is the flexibility to investigate a variety of statistics of interest through a carefully structured process.
Future work will focus on understanding the robustness of alternative measures that can be employed for algorithm assessment through meta-games, and developing effective tools for statistical analysis.

\section*{Acknowledgment}
This work was supported in part by funding from the US Army Research Office (MURI grant W911NF-18-1-0208), and a grant from Open Philanthropy.

\bibliographystyle{named}
\bibliography{bib}

\newpage
\appendix
\onecolumn
\section{Max-Entropy Nash}\label{appendix:max-ent}
In practice, it might be expensive to directly optimize the objective, due to the nonlinearity of the entropy function.
We instead optimize a piecewise linear approximation of the objective.
We have the following result.
\begin{theorem}
Given $\epsilon>0$, an $\epsilon$-maximum-entropy symmetric Nash can be solved by a mixed-integer linear program based on~\eqref{prog: mip} with an additional  $O(\lnorm{\hat{\itPi}}^2 / \epsilon)$
linear constraints.
\label{thm: max_ent}
\end{theorem}
\subsection{Proof of Theorem~\ref{thm: max_ent}}\label{sec: proof_mip}
\begin{proof}
The idea is to transform ~(\ref{prog: mip}) into the following mixed-integer linear program:
\begin{equation}
\begin{aligned}
\min_{\sigma^*} \quad & \sum_{\pi\in \hat{\itPi}}\gamma_{\pi}\\
\textrm{s.t.} \quad
& \forall \pi\in \hat{\itPi},\\
& \forall k\in[K], \gamma_{\pi}\ge l_k(\sigma^*(\pi))\\
\quad & 
    u_{\pi}= \sum_{\pi^\prime\in\hat{\itPi}}\sigma^*(\pi)u(\pi, \bm{\pi^\prime})\\
    & u^*\ge u_{\pi}\\
    &u^*-u_{\pi}\le Ub_{\pi}\\
    &\sigma(\pi)\le 1-b_{\pi}\\
  &\sigma^*(\pi)\geq0 \\
  &\sum_{\pi}\sigma^*(\pi)=1\\
&b_{\pi}\in\{0, 1\}\\
\end{aligned}
\label{prog: mip-approx}
\end{equation}

Where $l_k(x)=f(\frac{k}{K})+\frac{f(\frac{k+1}{K})-f(\frac{k}{K})}{\frac{1}{K}}(x-f(\frac{k}{K}))$ is the $k$-th piecewise linear segment of the function $f(x)=x\log x$, for a total of $K$ segments.
Denote the convex envelope of all these $K$ segments as $l(x)$.
The additional constraints are $\gamma_{\pi}\ge l_k(\sigma^*(\pi))$.
When we solve~\eqref{prog: mip-approx} exactly, one of these $K$ inequalities will be satisfied at equality.

Let $f(x^*)$ be the minimum of $f$ and $l(x^\prime)$ be the minimum of the piecewise approximation.
Then if $\lnorm{f-l}_{\infty}<\epsilon$, we can see that $f(x^*)\ge l(x^{*})-\epsilon\ge l(x^{\prime})-\epsilon$.
Then $x^\prime$ is an $\epsilon$-optimal solution of $f$.

Then to achieve an $\epsilon$-optimal solution of (\ref{prog: mip}) we need to choose some $K$, such that $\lnorm{f-l}_{\infty}<\frac{\epsilon}{\lnorm{\hat{\itPi}}}$.

For the segment $I_k=[\frac{k}{m}, \frac{k+1}{m}]$, by elementary calculus we can find $g(k)=\max_{x\in I_k}\lnorm{x\log x-l_k(x)}=\frac{k+1}{eK}(1+\frac{1}{k})^k-\frac{k(k+1)}{K}\log(1+\frac{1}{k})$.
This happens when $x=\frac{k+1}{eK}\left(1+\frac{1}{k}\right)^k$.

We will now prove
\begin{align*}
    g^\prime(k)&=\frac{e+\left(\left(1+\frac{1}{k}\right)^k(1+k)-e(1+2k)\right)\log\left(1+\frac{1}{k}\right)}{eK}<0\\
\end{align*}

To show this, by arranging the numerator, we need to show that 
\begin{align*}
    \left(k+(1+k)\left(1-\frac{1}{e}\left(1+\frac{1}{k}\right)^k\right)\right)\log\left(1+\frac{1}{k}\right)>1
\end{align*}

First notice that
\begin{align*}
    1-\frac{1}{e}\left(1+\frac{1}{k}\right)^k&=1-e^{k\log(1+\frac{1}{k})-1}\\
    &=1-e^{-\frac{1}{2k}+\frac{1}{3k^2}-\frac{1}{4k^3}+\frac{1}{5k^4}+\dotsc}\\
    &\ge 1-e^{-\frac{1}{2k}+\frac{1}{3k^2}}\\
    &\ge (-\frac{1}{2k}+\frac{1}{3k^2})-\frac{1}{2}\left(-\frac{1}{2k}+\frac{1}{3k^2}\right)^2\\
    &=\frac{1}{2k}-\frac{11}{24k^2}+\frac{1}{6k^3}-\frac{1}{18k^4}
\end{align*}
Then when $k\ge7$
\begin{align*}
    &(1+k)\left(1-\frac{1}{e}\left(1+\frac{1}{k}\right)^k\right)\\
    \ge&(1+k)\left(\frac{1}{2k}-\frac{11}{24k^2}+\frac{1}{6k^3}-\frac{1}{18k^4}\right)\\
    =&\frac{1}{2}+\frac{1}{24k}-\frac{7}{24k^2}+\frac{1}{9k^3}-\frac{1}{18k^4}>\frac{1}{2} 
\end{align*}
Therefore 
\begin{align*}
    \left(k+(1+k)\left(1-\frac{1}{e}\left(1+\frac{1}{k}\right)^k\right)\right)\log\left(1+\frac{1}{k}\right)>\left(k+\frac{1}{2}\right)\log\left(1+\frac{1}{k}\right)
\end{align*}

Let $h(k)=\left(k+\frac{1}{2}\right)\log\left(1+\frac{1}{k}\right)$.
Then $h^\prime(k)=\log\left(1+\frac{1}{k}\right)-\frac{1}{2k}-\frac{1}{2(k+1)}$
$h^{\prime\prime}(k)=-\frac{1}{k(k+1)}+\frac{1}{2k^2}+\frac{1}{2(k+1)^2}>0$.
Therefore $h^\prime(k)$ increases.
And since $\lim_{k\rightarrow\infty}h^\prime(k)=0$ we can deduce $h^\prime(k)<0$.
Therefore $h(k)$ is decreasing.
And since $\lim_{k\rightarrow\infty}h(k)=1$ we can see that $h(k)>1$.
Therefore the we have proved $g^\prime(k)<0$ for $k\ge7$ is decreasing.
It is easy to verify it also holds for $k<7$.
So $g(k)$ achieves maximum $\frac{1}{eK}$ when $k=0$.

Then we have when $K=\left\lfloor\frac{\lnorm{\hat{\itPi}}}{e\epsilon}\right\rfloor+1$, we can obtain an $\epsilon$-optimal maximum entropy Nash by solving (\ref{prog: mip-approx}) with additional $K\cdot (\lnorm{\itPi})=O\left(\frac{\lnorm{\hat{\itPi}}^2}{\epsilon}\right)$ linear constraints.

\end{proof}

\subsection{Setup}
We use GUROBI \citep{gurobi} as the solver.
In our experiments we always solve for a 0.05-optimal max-entropy Nash.

\section{Details of Gumbel IS-MCTS}
\label{sec: detail-g-search}
\subsection{Value Estimation}\label{sec: qhat}
When $C(s, a) > 0$, we simply let $\hat{q}(s, a)=\frac{R(s, a)}{C(s, a)}$.
When $C(s, a)=0$, following \citep{danihelka2022policy}, we let
$$\hat{q}(s, a)=\frac{1}{1+\sum_bC(s,b)}\left(\bm{v}(s,a)+\frac{\sum_bC(s,b)}{\sum_{b:C(s,b)>0}\bm{p}(s, b)}\sum_{b:C(s, b)>0}\frac{R(s, b)}{C(s, b)}\bm{p}(s, b)\right)$$ as an estimator.
And we use $G(\hat{q}(s, a))=c_2(c_1+\max_bC(s, b))\hat{q}(s,a)$, for some $c_1, c_2>0$.

\subsection{Action Selection at Non-Root Nodes}
\label{sec: non-root}
At a non-root node, an action is selected to minimize the discrepancy between 
$\Improve(\bm{p})$ and the produced visited frequency:
\begin{equation}
\arg\min_a\sum_b\left(\Improve(\bm{p})(s_i, b)-\frac{C(s_i, b)+1\{a=b\}}{1+\sum_cC(s_i, c)}\right)^2 \label{eq: non-root}
\end{equation}
\section{Algorithms Pseudocode}\label{sec:pseudocode}

\subsection{Vanilla AlphaZero Search}
Pseudocode presented as Alg.~\ref{algorithm: va-mcts}.
\begin{algorithm}[ht]
   \caption{Vanilla AlphaZero-styled IS-MCTS}
\begin{algorithmic}[1]
\FUNCTION{\texttt{VA-Search}($s$, $\bm{v}, \bm{p}$)}
   \STATE $\forall (s, a), R(s, a)=0, C(s, a)=0.$\;
   \STATE $\forall a\in\cA(s)$, sample $d(a)\sim Dirichlet(\alpha)$ 
   \FOR{$iter=1,\dotsc,num\_sim$}
   \STATE Sample a world state: $h \sim \Pr(h\mid s, \bm{p})$
   \WHILE{}
   \IF{$h$ is terminal}
   \STATE $\bm{r}\gets$ payoffs of players. {\bf Break}
   \ELSIF{$i\triangleq\tau(h)$ is chance}
   \STATE $a \gets$ sample according to chance
   \ELSIF{$s_{i}(h)$ not in search tree}
   \STATE Add $s_{i}(h)$ to search tree.
   \STATE $\bm{r}\leftarrow \bm{v}(s_{i}(h))$. {\bf Break}
   \ELSIF{$s_{i}(h)$ is root node $s$}
   \STATE $a\gets \arg\max_a \frac{R(s_i, a)}{C(s_i, a)}+c_{puct}((1-\epsilon)\bm{p}(s, a)+\epsilon d(a))\frac{\sqrt{C(s_i, a)}}{1+\sum_bC(s_i, b)}$ .
   \ELSE
   \STATE $a\gets \arg\max_a \frac{R(s_i, a)}{C(s_i, a)}+c_{puct}\bm{p}(s, a)\frac{\sqrt{C(s_i, a)}}{1+\sum_bC(s_i, b)}$
   \ENDIF
   \STATE $h\leftarrow ha$
   \ENDWHILE
   \FOR{$(s_{i}, a)$ in this trajectory}
   \STATE $R(s_i, a) \mathrel{+}= \bm{r}_i, C(s_i, a)  \mathrel{+}= 1$
   \ENDFOR
   \ENDFOR
   \STATE {\bf return} Action $a=\argmax C(s, a)$, (During training) a policy target that is the normalized visited count of $s$.
 \ENDFUNCTION
 \label{algorithm: va-mcts}
\end{algorithmic}
\end{algorithm}

\subsection{Sequential Halving}
Pseudocode presented as Alg.~\ref{algorithm: sh}.
\begin{algorithm}[h]
   \caption{Sequential Halving}
\begin{algorithmic}[1]
\FUNCTION{\texttt{Seq-Hal}($\hat{\cA}, K$)}
    \STATE Maintain a static variable $epoch$ across different calls of this function, initialized as 0\;
    \IF{$epoch==0$}
    \STATE $\hat{\cA}\leftarrow$ $K$ actions with largest $g(a)+\text{logit }\bm{p}(s, a)$\;
    \STATE $epoch\mathrel{+}=1$\;
    \ENDIF
    \STATE $a\gets$ an action that has not been visited $\lfloor\frac{sim\_num}{\frac{K}{2^{epoch-1}}\log_2 K}\rfloor$ times at current epoch in $\hat{\cA}$\;
    \IF{All actions in $\hat{\cA}$ have been visited $\lfloor\frac{sim\_num}{\frac{K}{2^{epoch-1}}\log_2 K}\rfloor$ times in current epoch}
    \STATE $\hat{\cA}\gets$ Top $\lfloor\frac{K}{2^{epoch}}\rfloor$ actions in $\hat{\cA}$ based on $g(a)+\text{logit } \bm{p}(s, a)+G(\hat{q}(s, a))$\;
    \STATE $epoch\gets epoch+1$\;
    \ENDIF
   \STATE {\bf return} $a, \hat{A}$.
 \ENDFUNCTION
 \label{algorithm: sh}
\end{algorithmic}
\end{algorithm}

\subsection{Self-play based Training}
Pseudocode presented as Alg.~\ref{algorithm: self-play}.
\begin{algorithm}[h]
   \caption{Self-Play Training for Search Methods}
\begin{algorithmic}[1]
\FUNCTION{\texttt{Self-Play-Train}($\cG$)}
   \STATE Initialize $\bm{v}, \bm{v}^\prime, \bm{p}, \bm{p}^\prime$\;
   \STATE $D_{\bm{v}}=\{\}, D_{\bm{p}}=\{\}$
   \FOR{$iter=1,\dotsc,num\_epoch$}
   \STATE $h\leftarrow$ Initial state $h_0$ of $\cG$\;
   \WHILE{}
   \IF{$h$ is terminal}
   \STATE $\bm{r}\gets$ payoffs of players {\bf Break}
   \ELSIF{$i\triangleq\tau(h)$ is chance}
   \STATE $a\leftarrow$ sample according to chance\; 
   \ELSE
   \STATE $a, \pi\leftarrow\textsc{Search}(s_i(h), \bm{v}^\prime, \bm{p}^\prime)$\;
   \STATE $D_{\bm{p}}\leftarrow D_{\bm{p}} \cup \{(s_i(h), \pi)\}$, $\pi$ is the visiting frequency during search\;
   \ENDIF
   \STATE $h\leftarrow ha$
   
   \ENDWHILE
   \FOR{$s_{i}$ in this trajectory}
   \STATE $D_{\bm{v}}\leftarrow D_{\bm{v}} \cup \{s_i, \bm{r}_i\}$
   \ENDFOR
   \STATE $\bm{v}, \bm{p}\leftarrow \textsc{Update}(D_{\bm{v}}, D_{\bm{p}})$
   \STATE Replace parameters of $\bm{v}^\prime, \bm{p}^\prime$ with the latest parameters of $\bm{v}, \bm{p}$ periodically.
   \ENDFOR
   \STATE {\bf return} $\bm{v}, \bm{p}$
 \ENDFUNCTION
 \label{algorithm: self-play}
\end{algorithmic}
\end{algorithm}

\section{Hyperparameters}\label{sec:hyper}
\subsection{Input Representation}
We use the infostate\_tensor in OpenSpiel \citep{lanctot2019openspiel} as the representation of infostate.
In Bargaining games, this includes information (1) which player the current agent is playing (2) pool configuration (3) the player's own private valuation (4) the current round number.

\subsection{PPO Algorithms}
We use the same PPO hyperparameters for every place when it is used. 
See Table~\ref{tab:hyp-ppo}.
\begin{table*}[ht]
    \centering
    \begin{adjustbox}{max width=\textwidth}
    \begin{tabular}{l|l}
    {\bf Hyperparameter Name} & {\bf Values}\\
    \hline
    Learning rate  & $2e-4$  \\
    Optimizer & Adam\\
    Batch size & 16\\
    Number of minibatch & 4\\
    Number of updates per epoch & 10\\
    Number of steps per PPO trajectory & 64\\
    Number of game trajectories in total & 1e6\\
    Entropy weight & 0.01\\
    Clipped parameter $\epsilon$ & 0.2\\
    GAE lambda & 0.95\\
    RL discount factor & 1\\
    Torso for policy nets & [256, 256]\\
    Torso for value nets & [256, 256]\\
    \end{tabular}
    \end{adjustbox}
    \caption{Hyper-parameters for PPO.}
    \label{tab:hyp-ppo}
\end{table*}
For single-agent PPO and IDPPO, the PPO algorithm trains policy nets by minimizing the clipped loss $\mathbb{E}_t\left[\min\left(\frac{\pi_\theta(a_t\mid s_t)}{\pi_{old}(a_t\mid s_t)}\hat{A}_t, clip\left(\frac{\pi_\theta(a_t\mid s_t)}{\pi_{old}(a_t\mid s_t)}, 1+\epsilon, 1-\epsilon\right)\hat{A}_t\right)\right]$, where $A_t$ is estimated by \emph{generalized advantage estimation} (GAE) \citep{schulman2015high}.
Value nets are updated by minimizing L2 loss from the value targets produced by GAE.
For MAPPO, the loss is simply the summation of individual PPO losses of every player.

\subsection{R-NaD}
We fine-tuned the learning rate to be 1e-3, the total number of game trajectories to be around 1e6.
Others are the same as the default hyperparameters in OpenSpiel \citep{lanctot2019openspiel}.

\subsection{NFSP}
See Table~\ref{tab:hyp-nfsp}.
\begin{table*}[ht]
    \centering
    \begin{adjustbox}{max width=\textwidth}
    \begin{tabular}{l|l}
    {\bf Hyperparameter Name} & {\bf Values}\\
    \hline
    RL network Learning rate  & $0.01$  \\
    DQN buffer size & $2^{17}$\\
    SL network learning rate & $0.01$ \\
    Optimizer & Adam\\
    Batch size & 256\\
    Number of game trajectories in total & 1e6\\
    Torso for RL net & [256, 256]\\
    Torso for SL net & [256, 256]\\
    \end{tabular}
    \end{adjustbox}
    \caption{Hyper-parameters for NFSP.}
    \label{tab:hyp-nfsp}
\end{table*}

\subsection{PSRO}
See Table~\ref{tab:hyp-psro}.
\begin{table*}[ht]
    \centering
    \begin{adjustbox}{max width=\textwidth}
    \begin{tabular}{l|l}
    {\bf Hyperparameter Name} & {\bf Values}\\
    \hline
    Number of PSRO iterations  & $32$  \\
    best-response method & PPO\\
    Number of game trajectories to train each BR & 1e6\\
    Meta-strategy solver & Max-entropy Nash
    \end{tabular}
    \end{adjustbox}
    \caption{Hyper-parameters for PSRO.}
    \label{tab:hyp-psro}
\end{table*}
\subsection{FCP}
See Table~\ref{tab:hyp-fcp}
\begin{table*}[ht]
    \centering
    \begin{adjustbox}{max width=\textwidth}
    \begin{tabular}{l|l}
    {\bf Hyperparameter Name} & {\bf Values}\\
    \hline
    Number of self-play runs & $32$  \\
    Number of checkpoint strategies considered per run & $100$\\
    Strategies picked each run & The first one, the last one, and the one that just achieves half of the social welfare of the last one\\
    Self-play method & IDPPO\\
    best-response method & PPO\\
    \end{tabular}
    \end{adjustbox}
    \caption{Hyper-parameters for FCP.}
    \label{tab:hyp-fcp}
\end{table*}

\subsection{Gumbel Search}
See Table~\ref{tab:hyp-g-search}.
\begin{table*}[ht]
    \centering
    \begin{adjustbox}{max width=\textwidth}
    \begin{tabular}{l|l}
    {\bf Hyperparameter Name} & {\bf Values}\\
    \hline
    Learning rate  & $2e-4$  \\
    Optimizer & SGD\\
    Batch size & 256\\
    Max buffer size & $2^{17}$\\
    $c_1$ & 50\\
    $c_2$ & 0.1\\
    $sim\_num$ & 200\\
    $K$ & 16\\
    Network delayed period & 1000\\
    Number of game trajectories in total & 1e6\\
    Torso for PVN & [256, 256]\\
    \end{tabular}
    \end{adjustbox}
    \caption{Hyper-parameters for Gumbel Search.}
    \label{tab:hyp-g-search}
\end{table*}

\subsection{VA Search}
See Table~\ref{tab:hyp-va-search}.
\begin{table*}[ht]
    \centering
    \begin{adjustbox}{max width=\textwidth}
    \begin{tabular}{l|l}
    {\bf Hyperparameter Name} & {\bf Values}\\
    \hline
    Learning rate  & $1e-3$  \\
    Optimizer & SGD\\
    Batch size & 256\\
    Max buffer size & $2^{17}$\\
    $c_{puct}$ & 20\\
    $sim\_num$ & 200\\
    Dirichlet $\alpha$ & $\frac{1}{\lnorm{\cA}}\cdot\bm{1}$\\
    $\epsilon$ & 0.25\\
    Network delayed period & 1000\\
    Number of game trajectories in total & 1e6\\
    Torso for PVN & [256, 256]\\
    \end{tabular}
    \end{adjustbox}
    \caption{Hyper-parameters for VA Search.}
    \label{tab:hyp-va-search}
\end{table*}

\section{$\NashConv$ Results}\label{sec:nashconv}
All $\NashConv$s are estimated using PPO as a best response method.
The results are reported across three random seeds.

\subsection{$\Barg(10, 0, 1)$}
\begin{figure*}[t]
\begin{tabular}{ccc}
\includegraphics[width=0.32\textwidth]{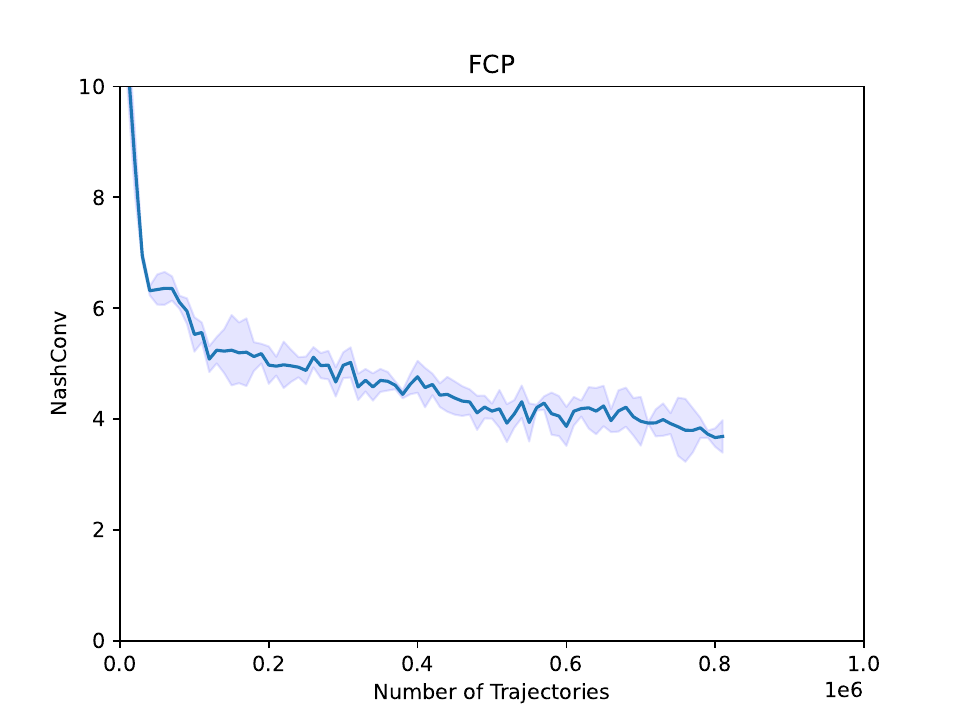} &
\includegraphics[width=0.32\textwidth]{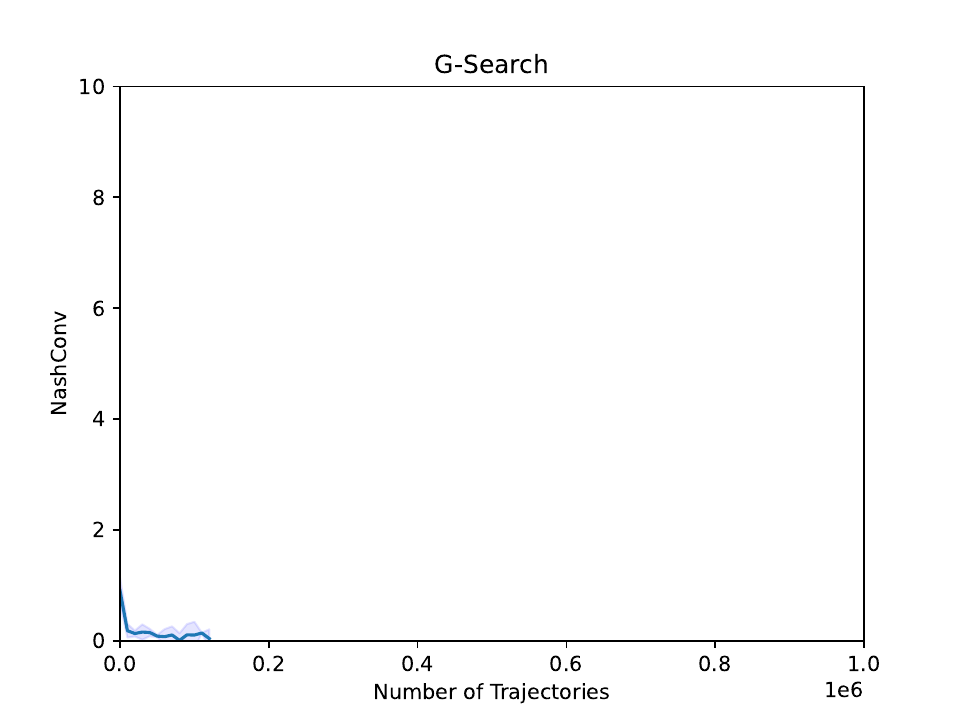} &
\includegraphics[width=0.32\textwidth]{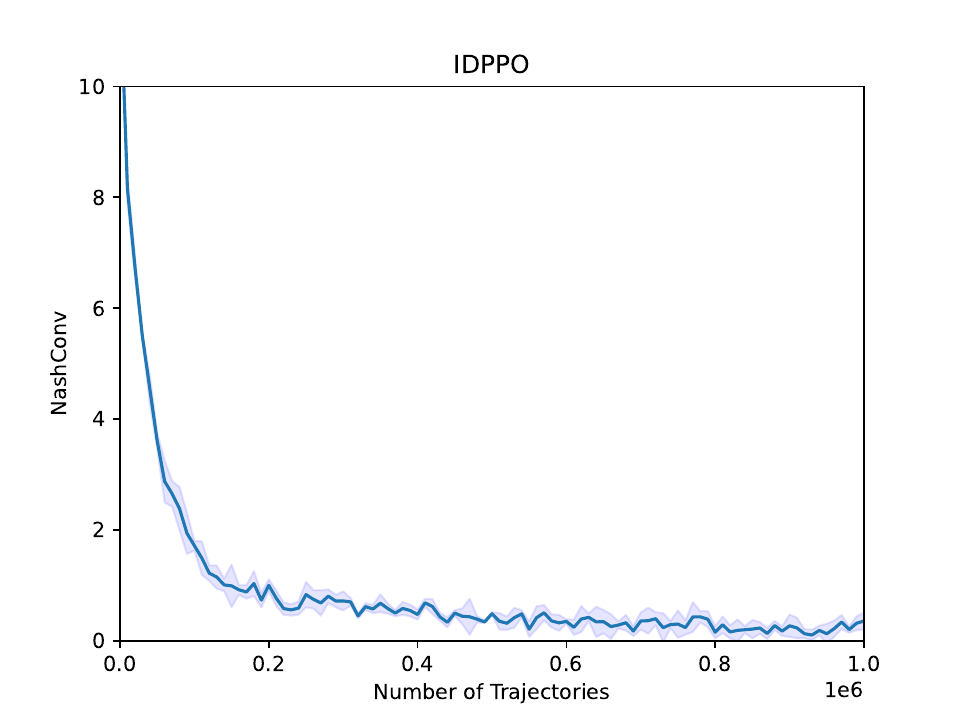}\\
\includegraphics[width=0.32\textwidth]{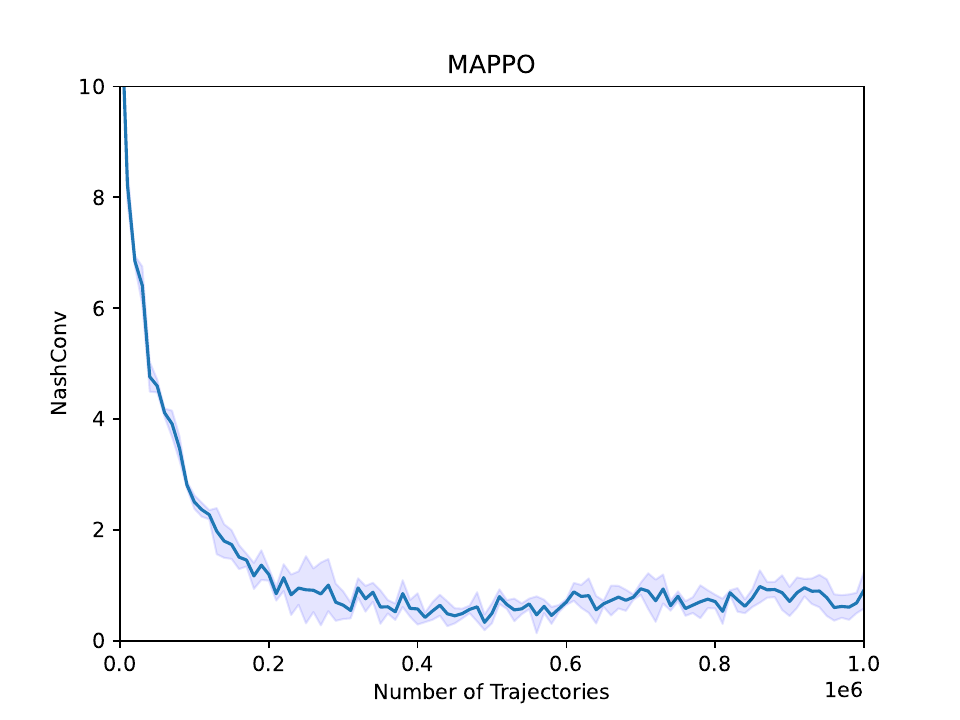} &
\includegraphics[width=0.32\textwidth]{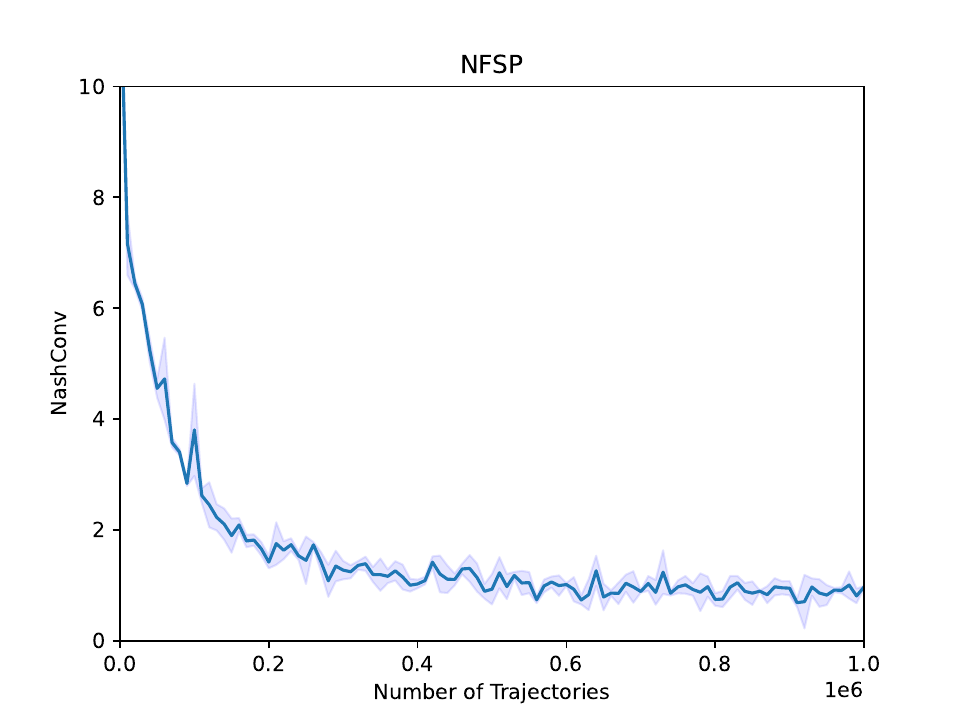} &
\includegraphics[width=0.32\textwidth]{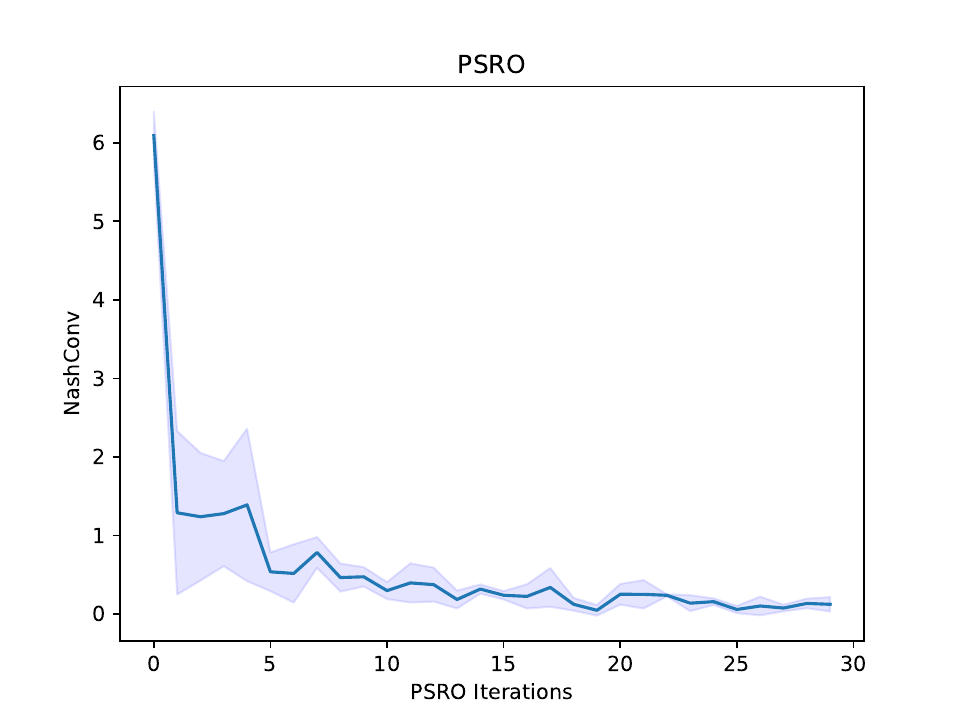}\\
\includegraphics[width=0.32\textwidth]{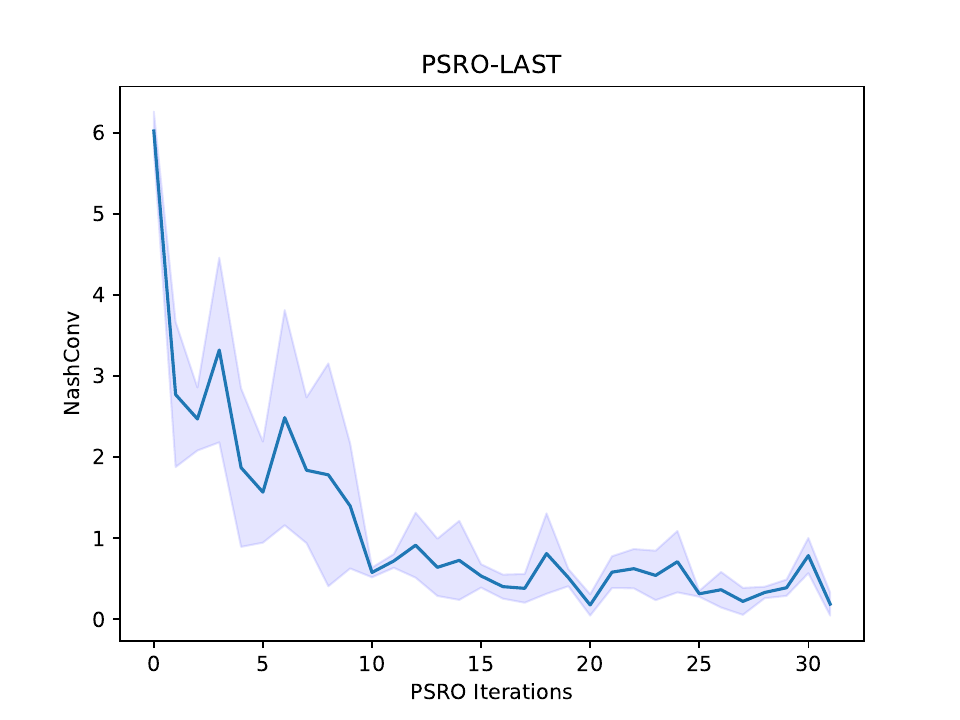} & 
\includegraphics[width=0.32\textwidth]{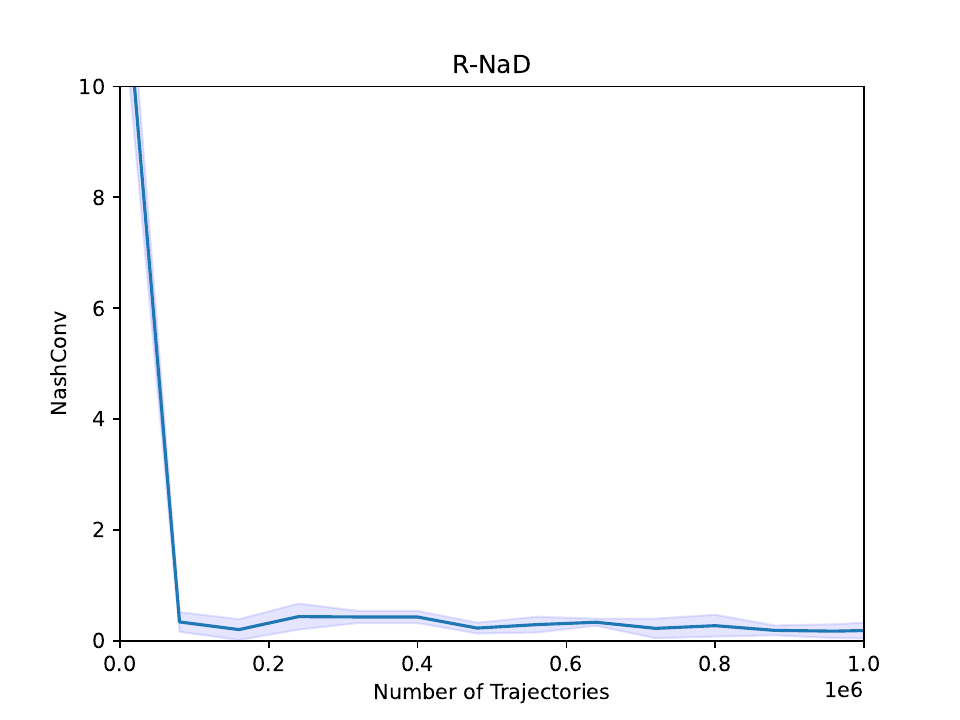} &
\includegraphics[width=0.32\textwidth]{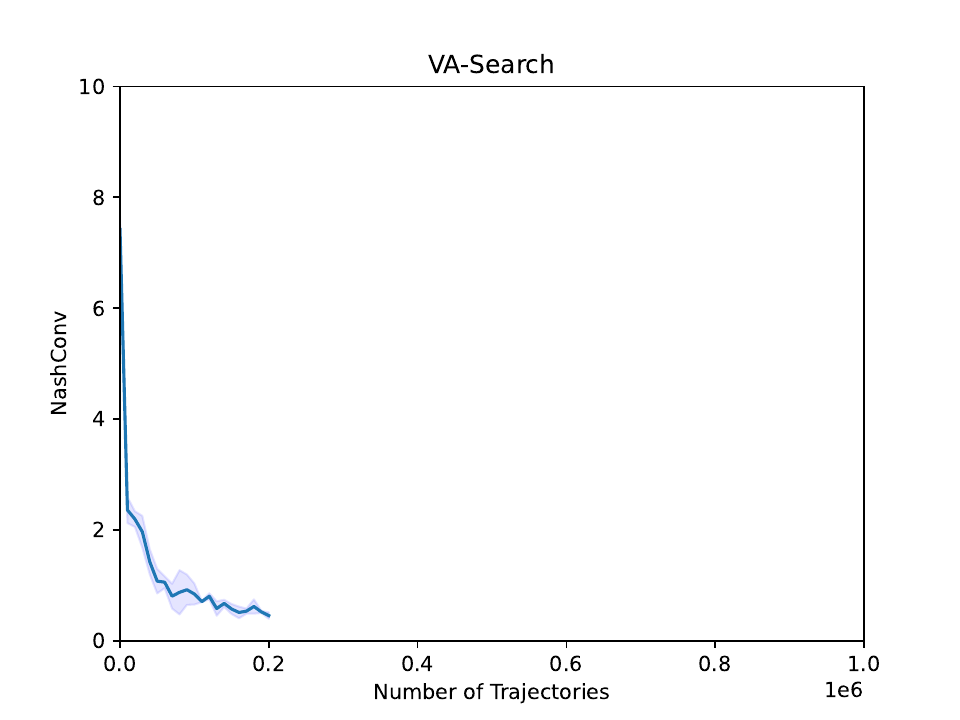}\\
\end{tabular}
\caption{$\NashConv$ of $\Barg(10, 0, 1)$}
\label{tab:nashconv-10}
\end{figure*}

See Figures~\ref{tab:nashconv-10}.

\subsection{$\Barg(30, 0.125, 0.935)$}
\begin{figure*}[t]
\begin{tabular}{ccc}
\includegraphics[width=0.32\textwidth]{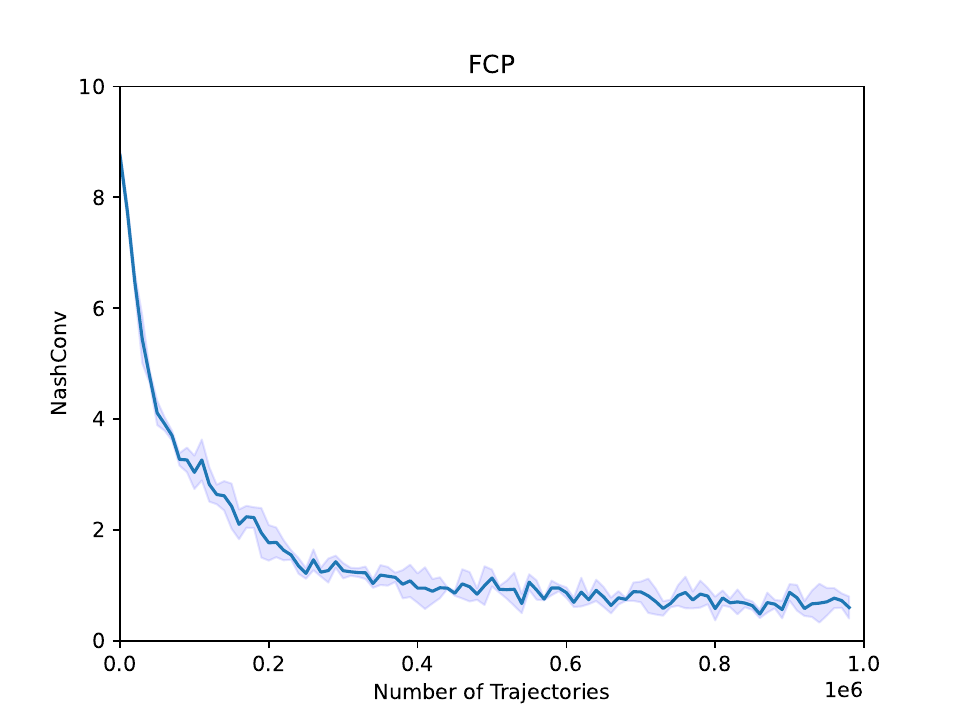} &
\includegraphics[width=0.32\textwidth]{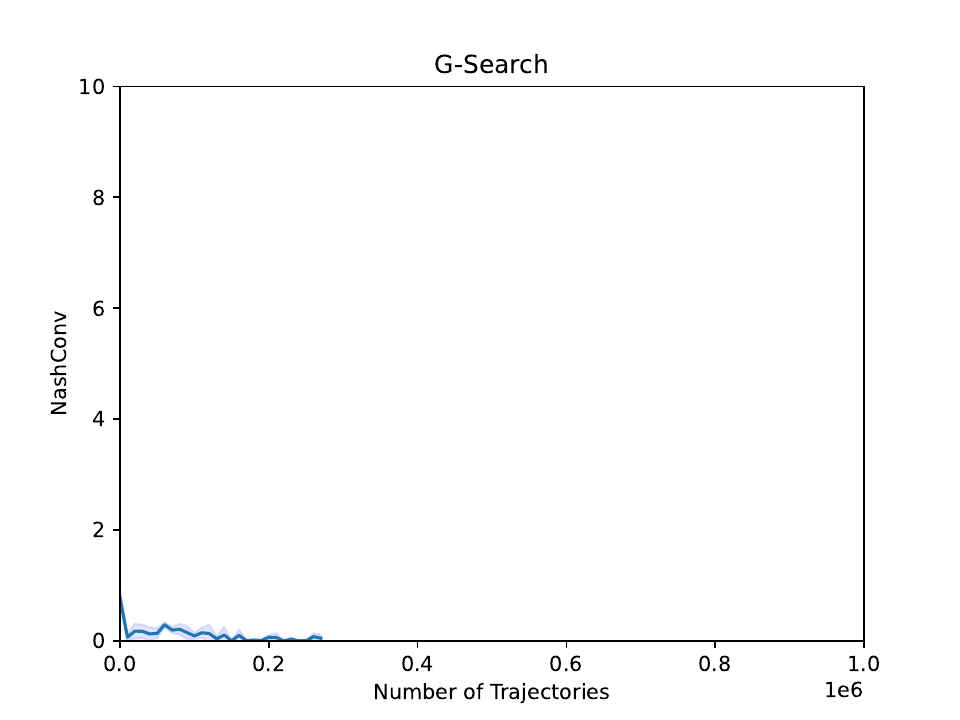} &
\includegraphics[width=0.32\textwidth]{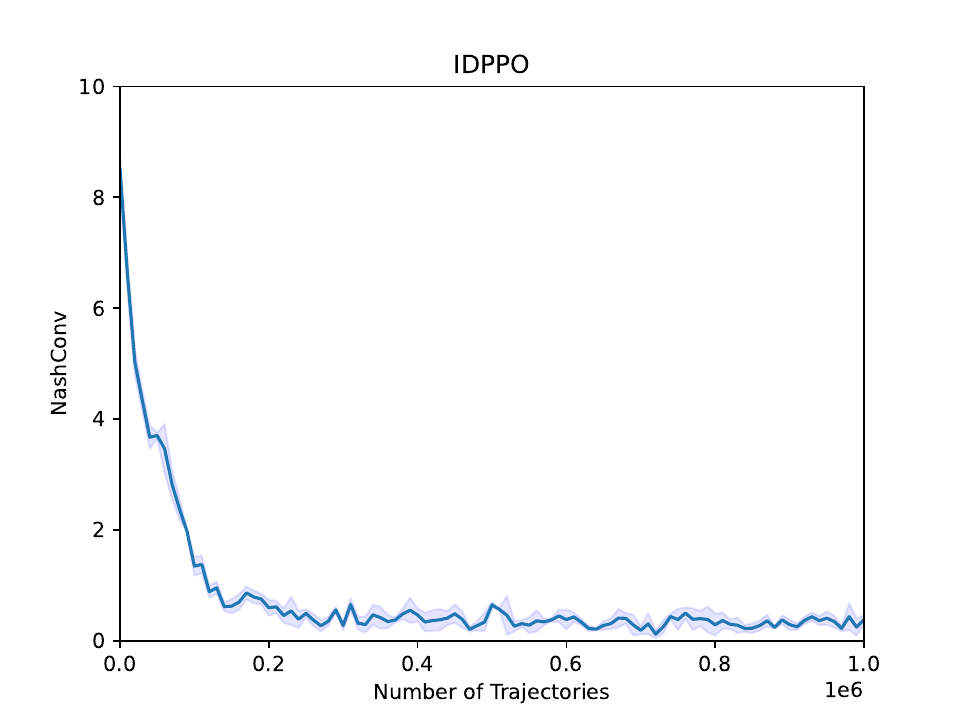}\\
\includegraphics[width=0.32\textwidth]{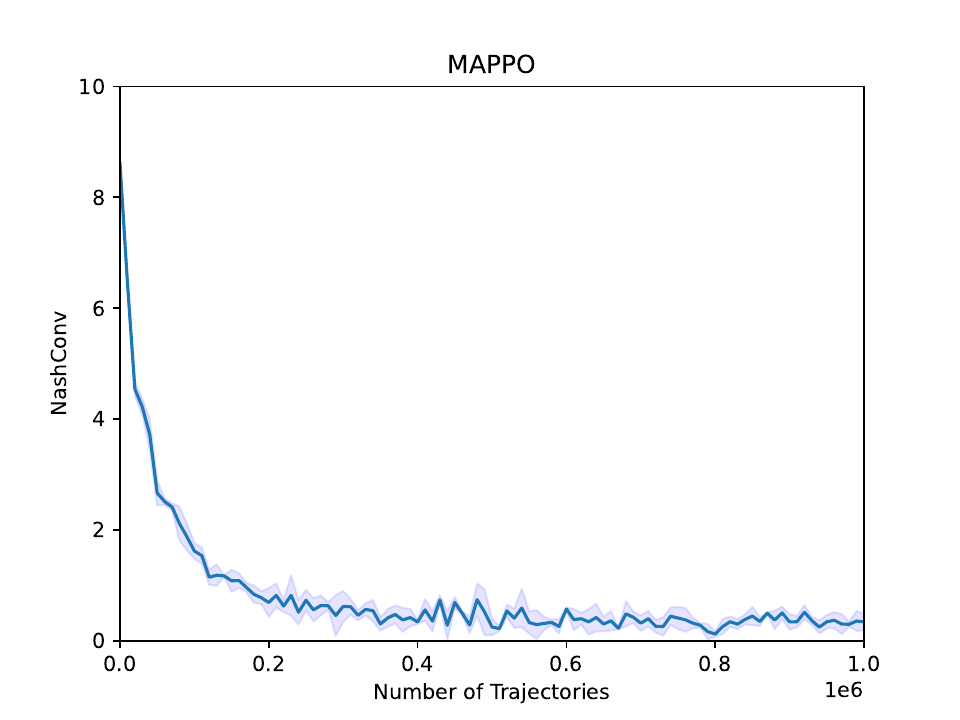} &
\includegraphics[width=0.32\textwidth]{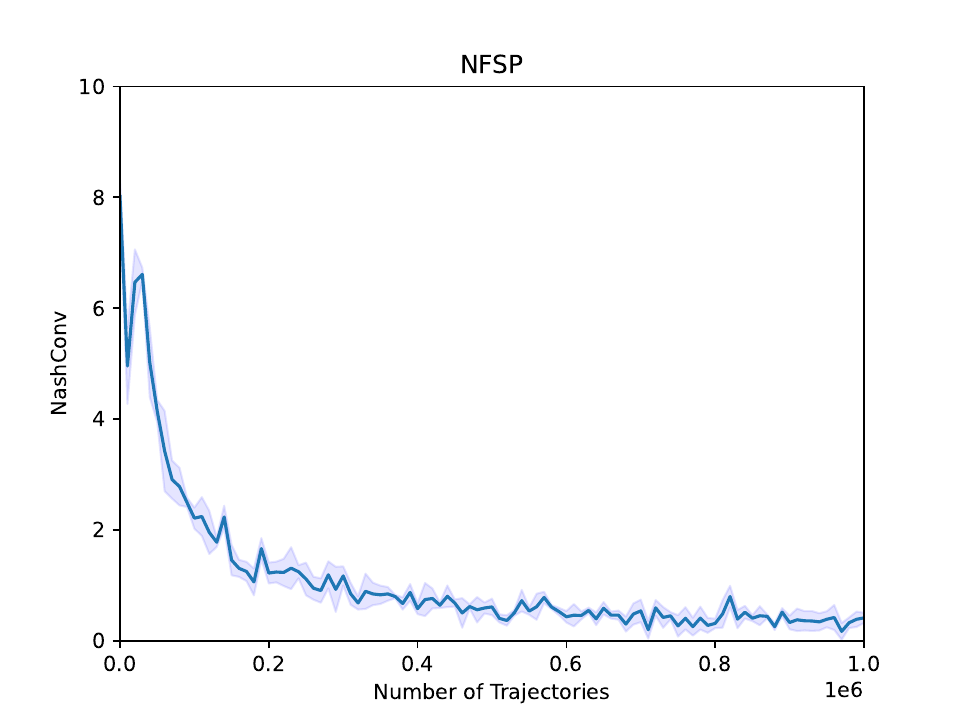} &
\includegraphics[width=0.32\textwidth]{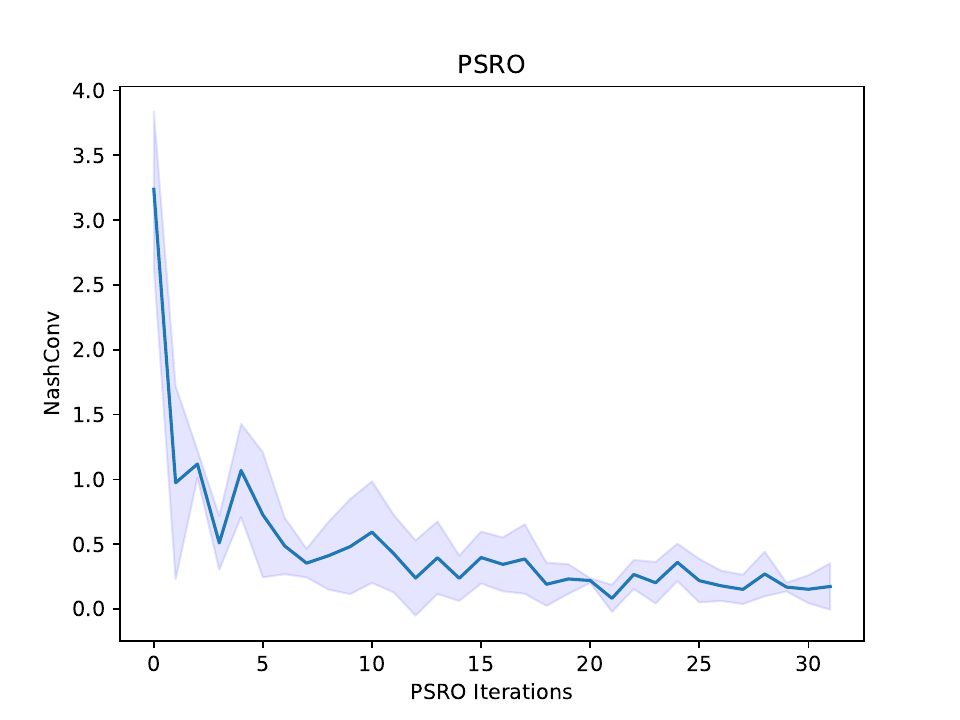}\\
\includegraphics[width=0.32\textwidth]{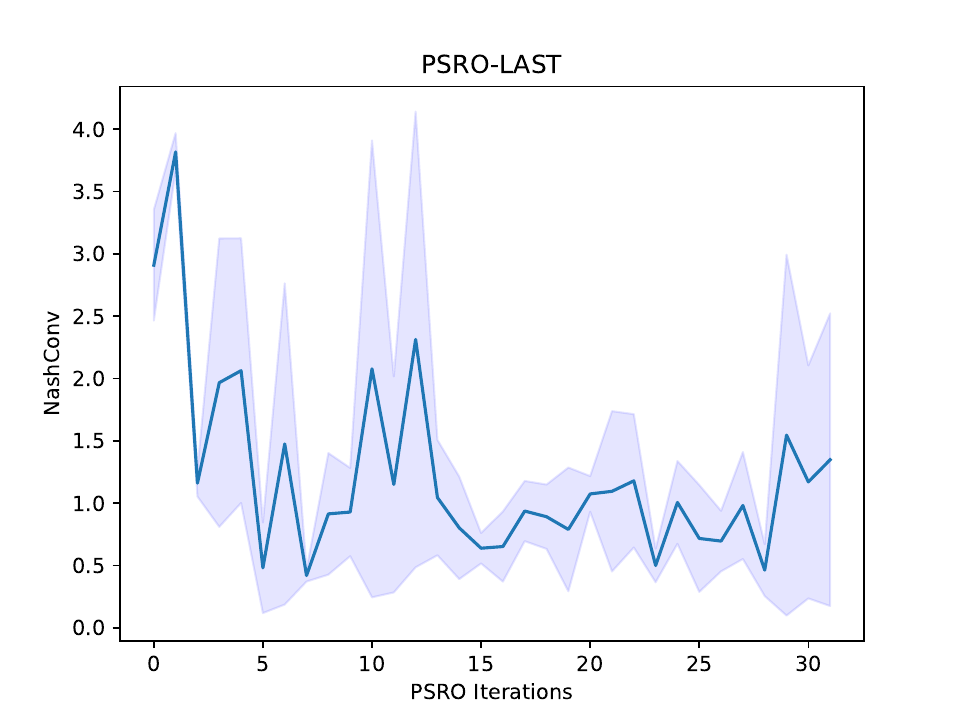} & 
\includegraphics[width=0.32\textwidth]{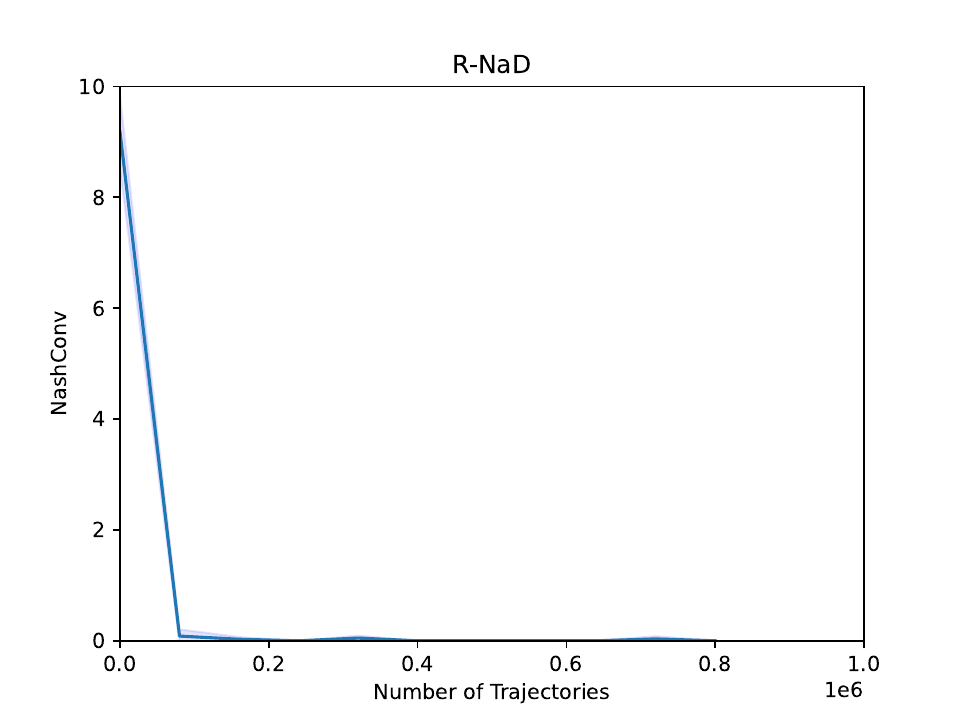} &
\includegraphics[width=0.32\textwidth]{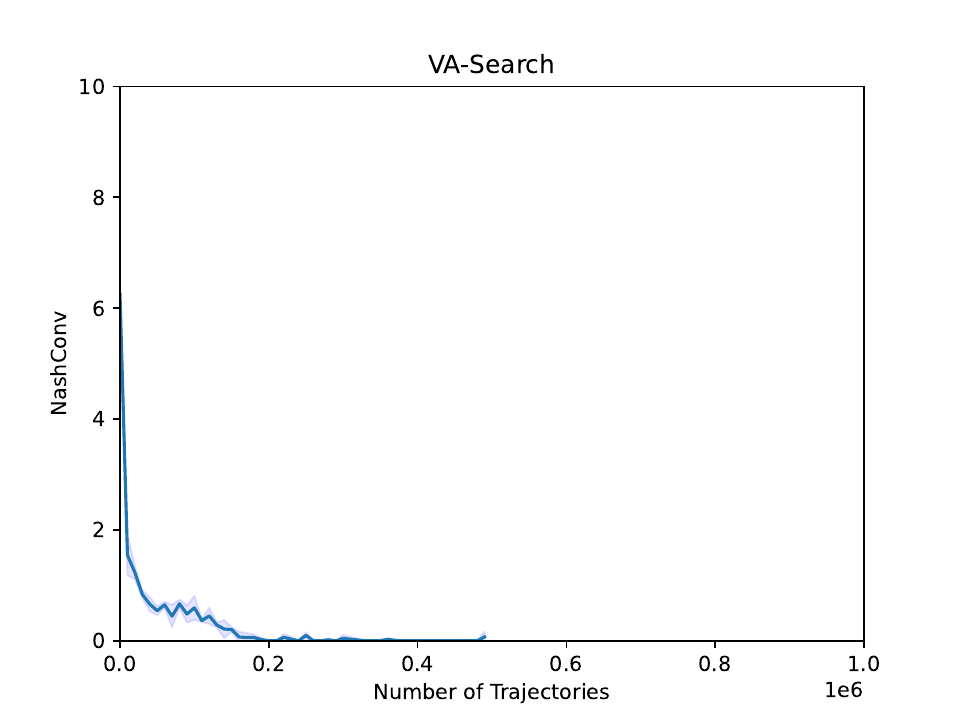}\\
\end{tabular}
\caption{$\NashConv$ of $\Barg(30, 0.125, 0.935)$}
\label{tab:nashconv-30}
\end{figure*}

See Figures~\ref{tab:nashconv-30}.

\section{Empirical Distribution of $\Regret$}\label{sec:regret-plots}
\subsection{$\Barg(10, 0, 1)$}
\begin{figure*}[t]
\begin{tabular}{cccc}
\includegraphics[width=0.24\textwidth]{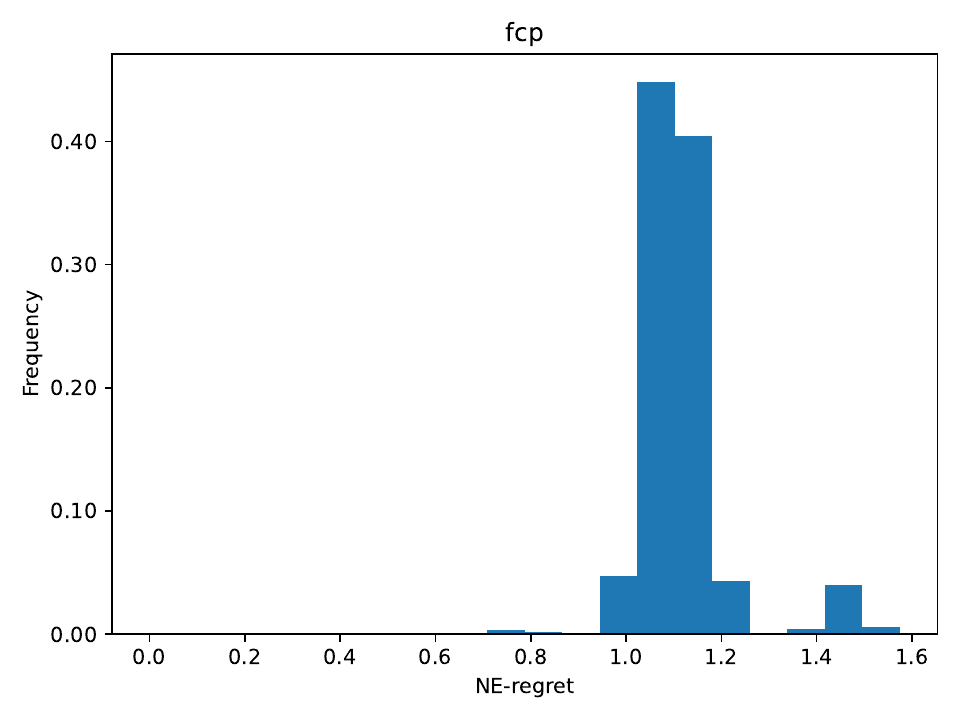} &
\includegraphics[width=0.24\textwidth]{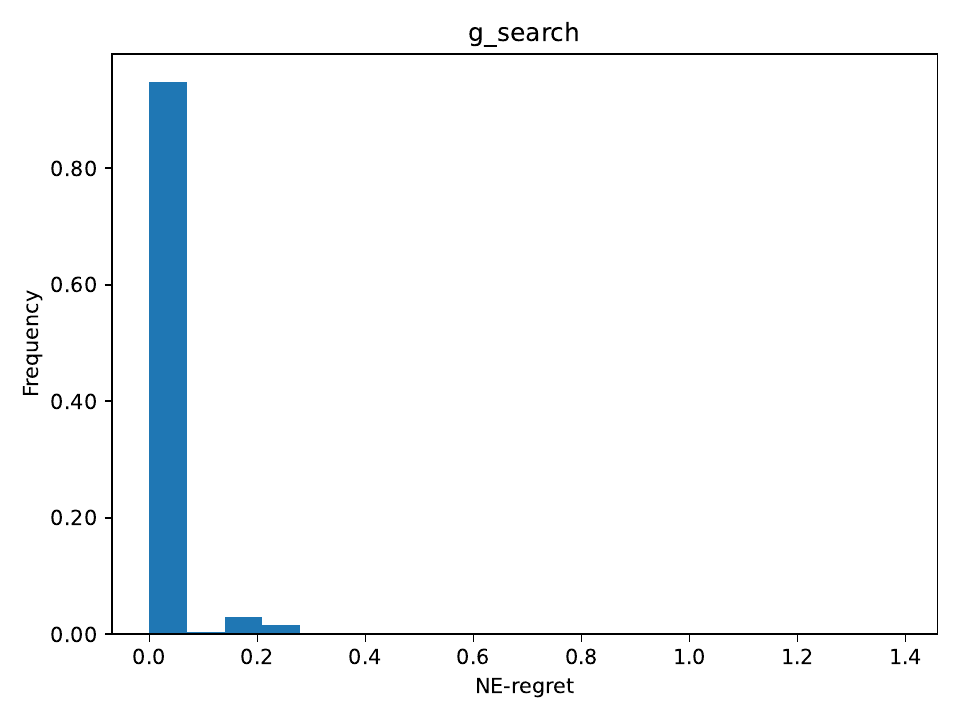} &
\includegraphics[width=0.24\textwidth]{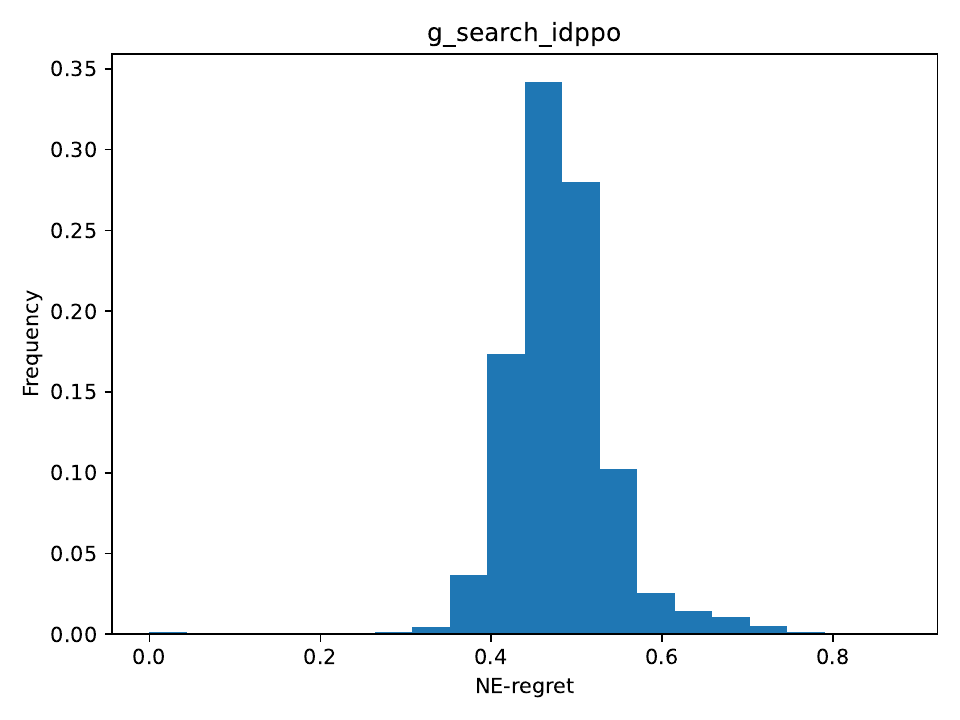} &
\includegraphics[width=0.24\textwidth]{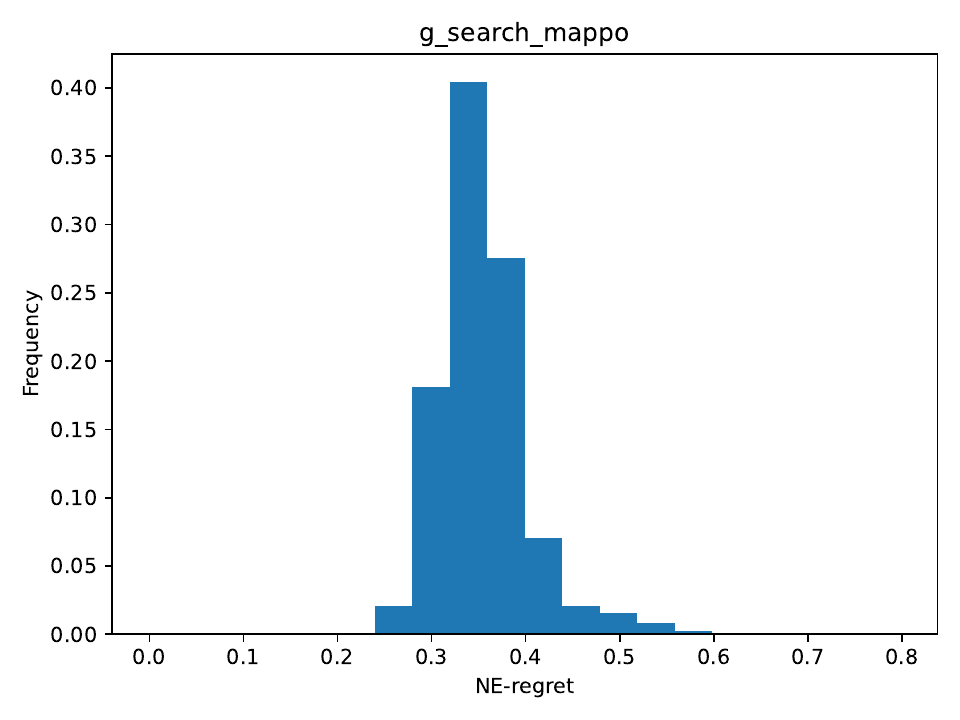} \\
\includegraphics[width=0.24\textwidth]{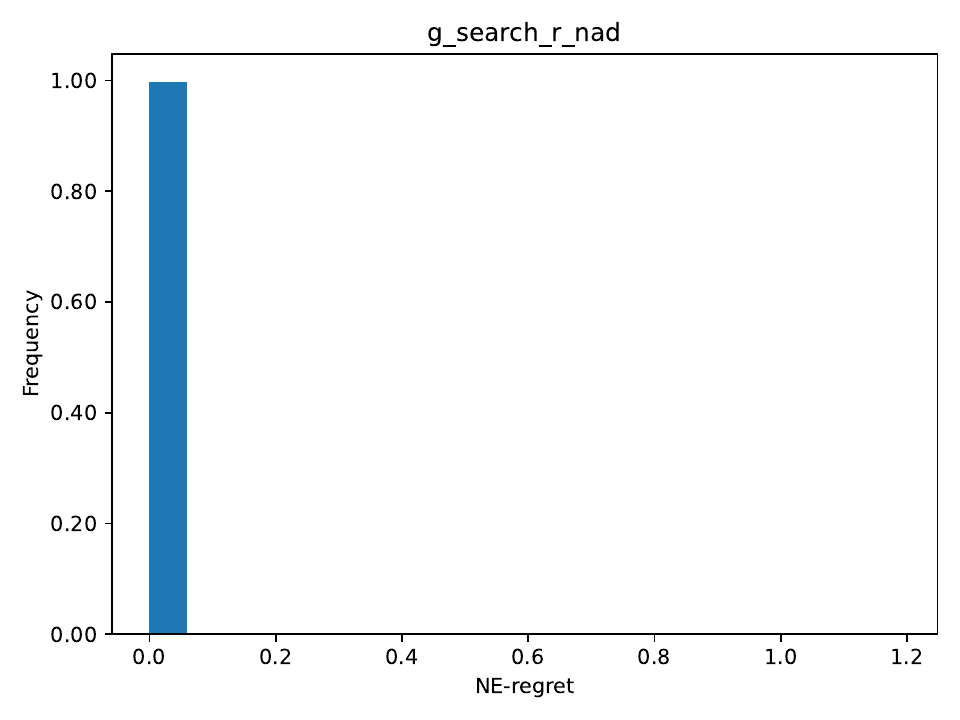} &
\includegraphics[width=0.24\textwidth]{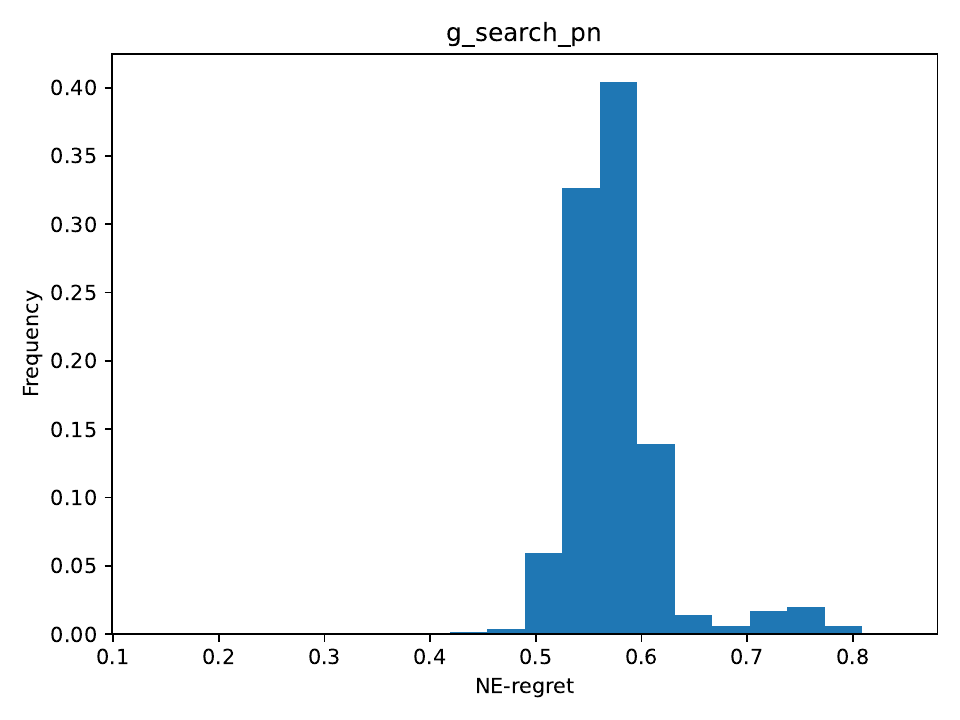} &
\includegraphics[width=0.24\textwidth]{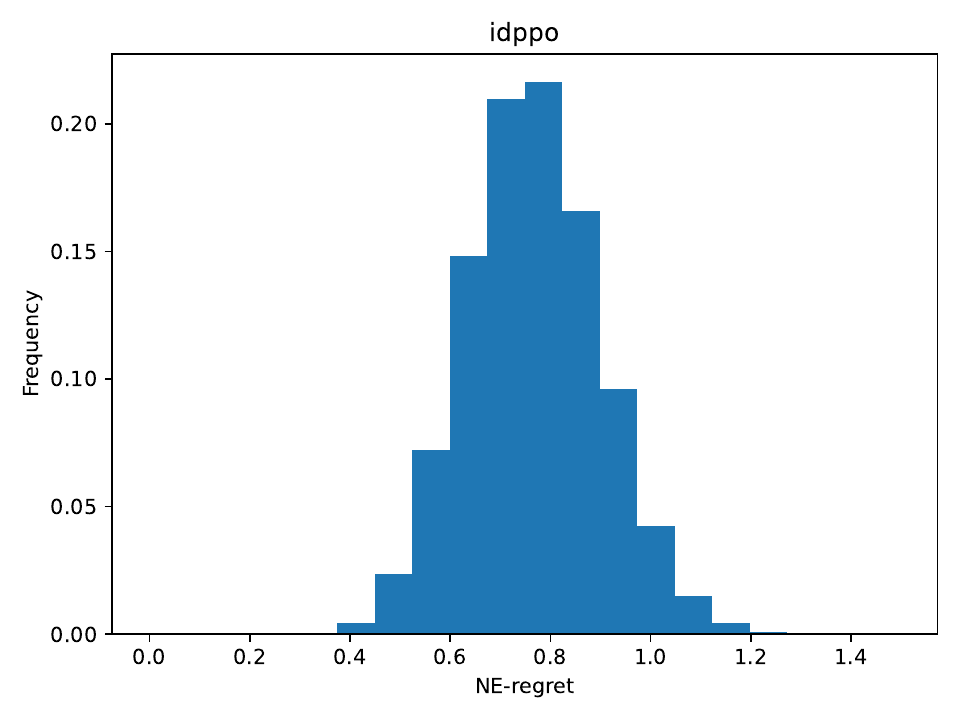} & 
\includegraphics[width=0.24\textwidth]{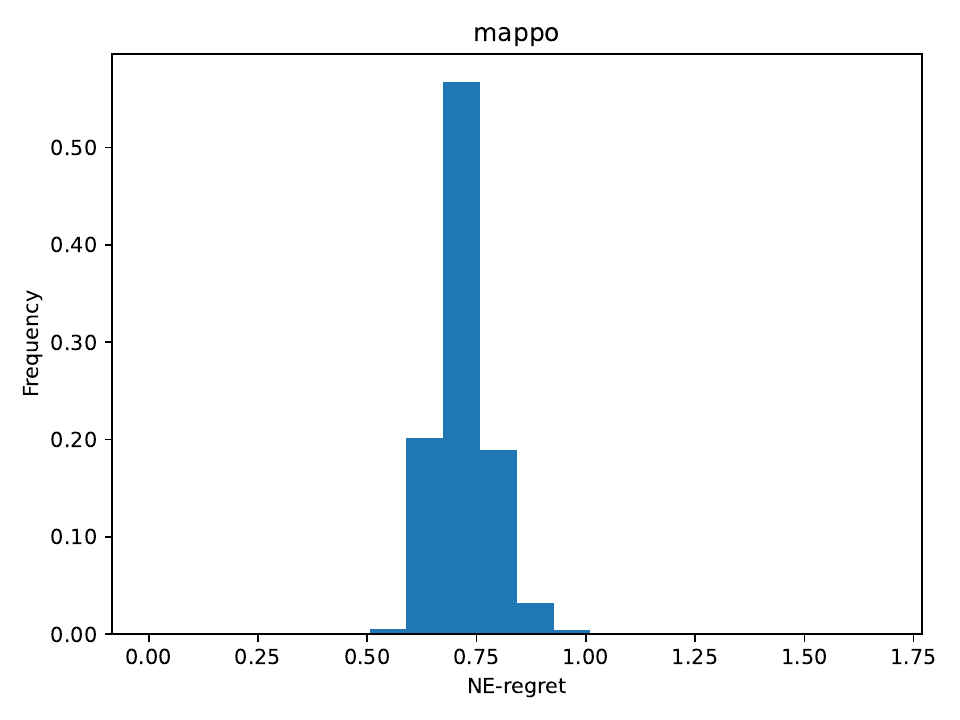}\\
\includegraphics[width=0.24\textwidth]{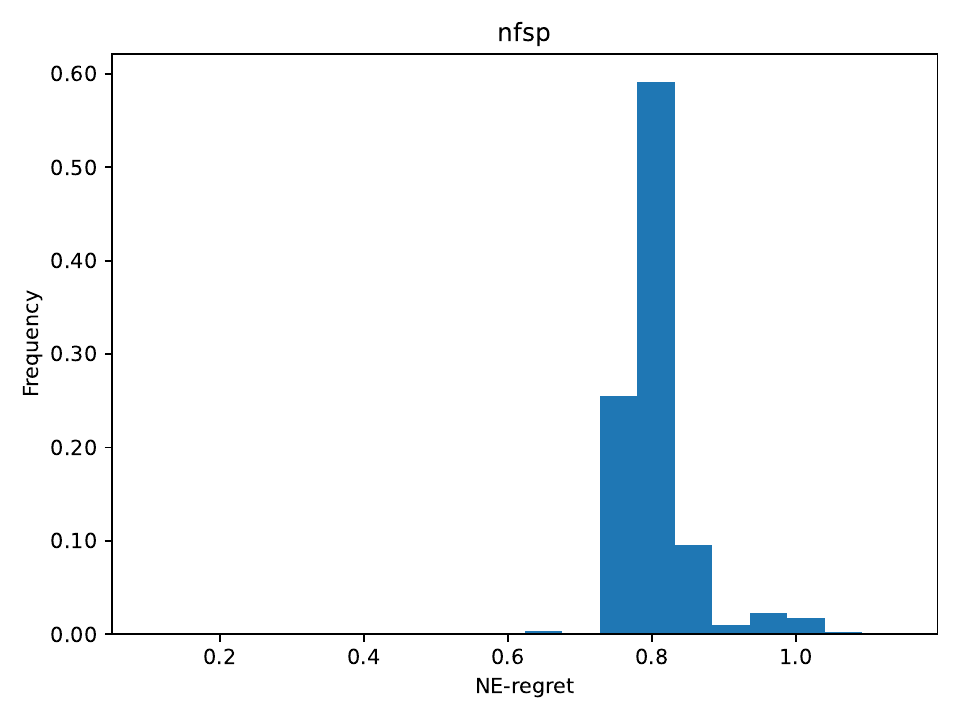} & 
\includegraphics[width=0.24\textwidth]{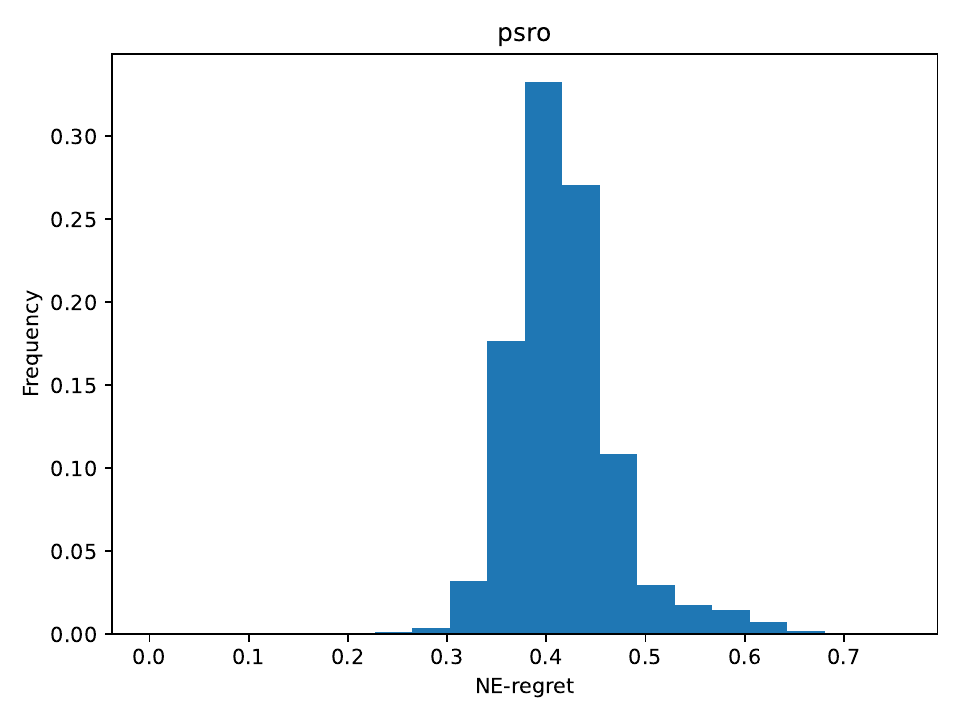} & 
\includegraphics[width=0.24\textwidth]{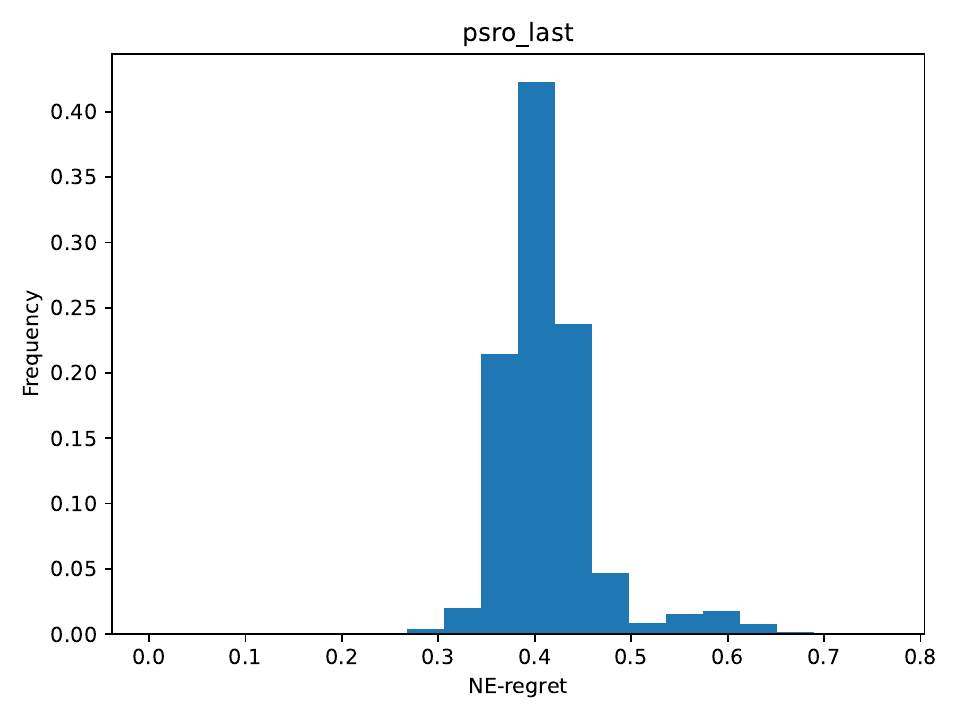} &
\includegraphics[width=0.24\textwidth]{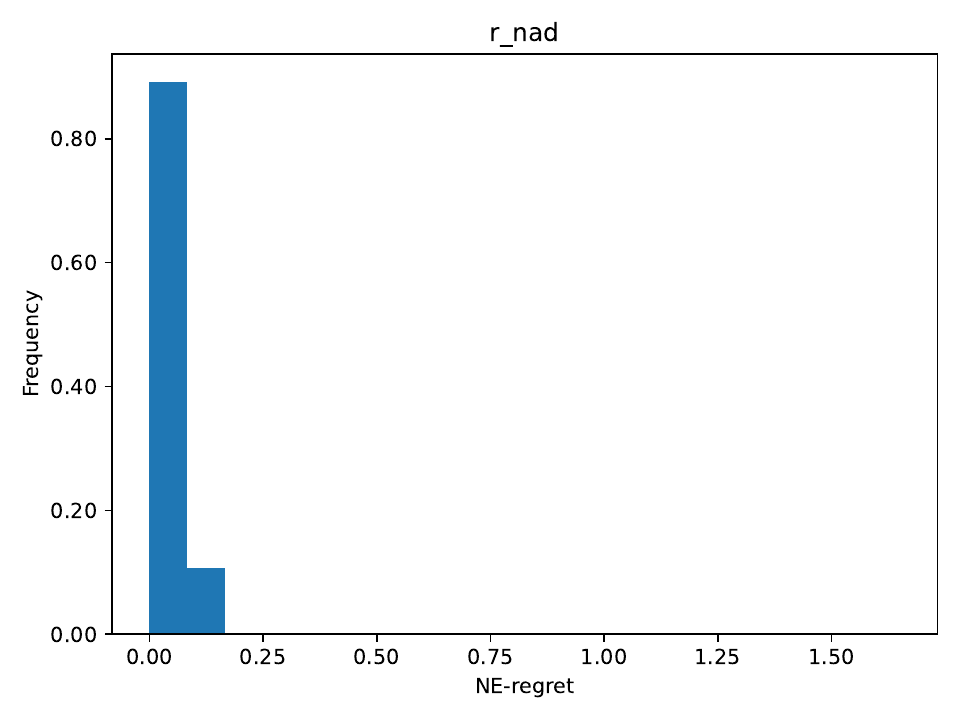}\\
\includegraphics[width=0.24\textwidth]{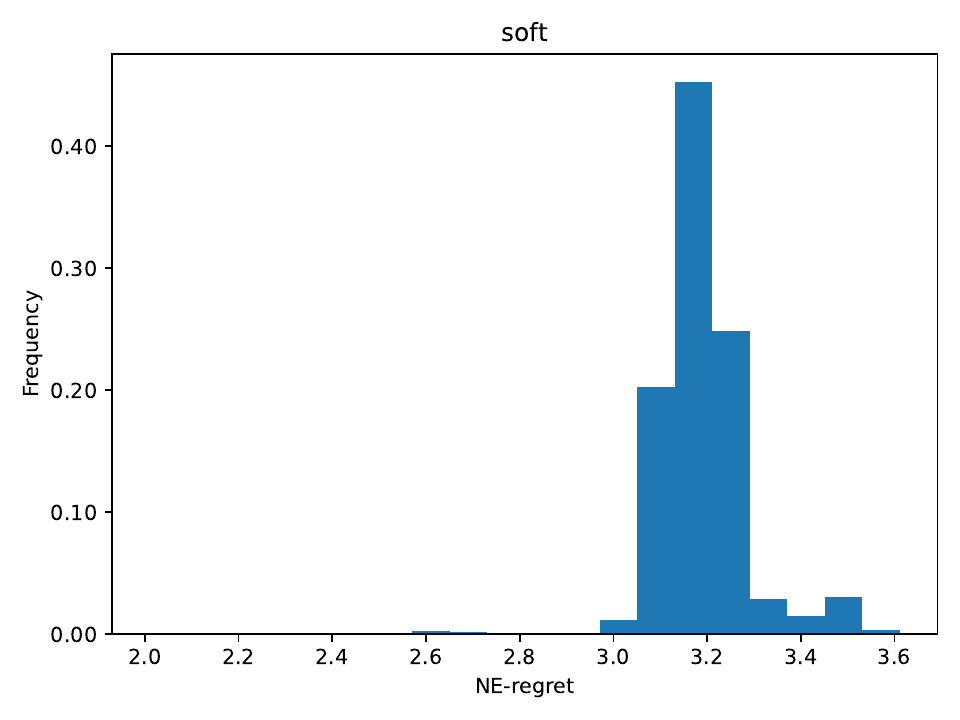} &
\includegraphics[width=0.24\textwidth]{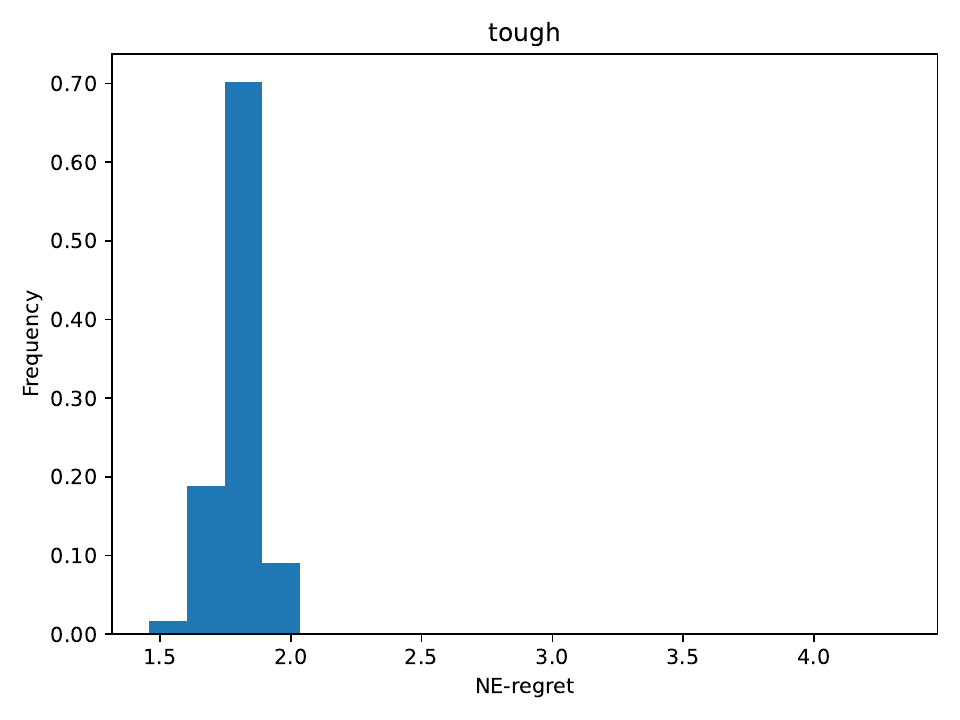} & 
\includegraphics[width=0.24\textwidth]{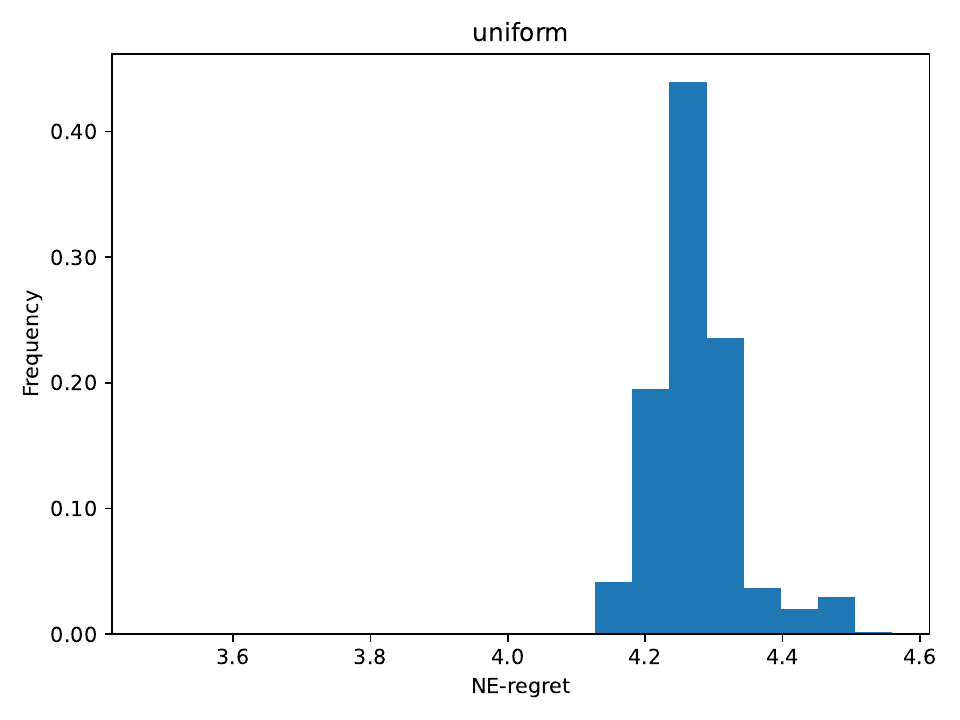} & 
\includegraphics[width=0.24\textwidth]{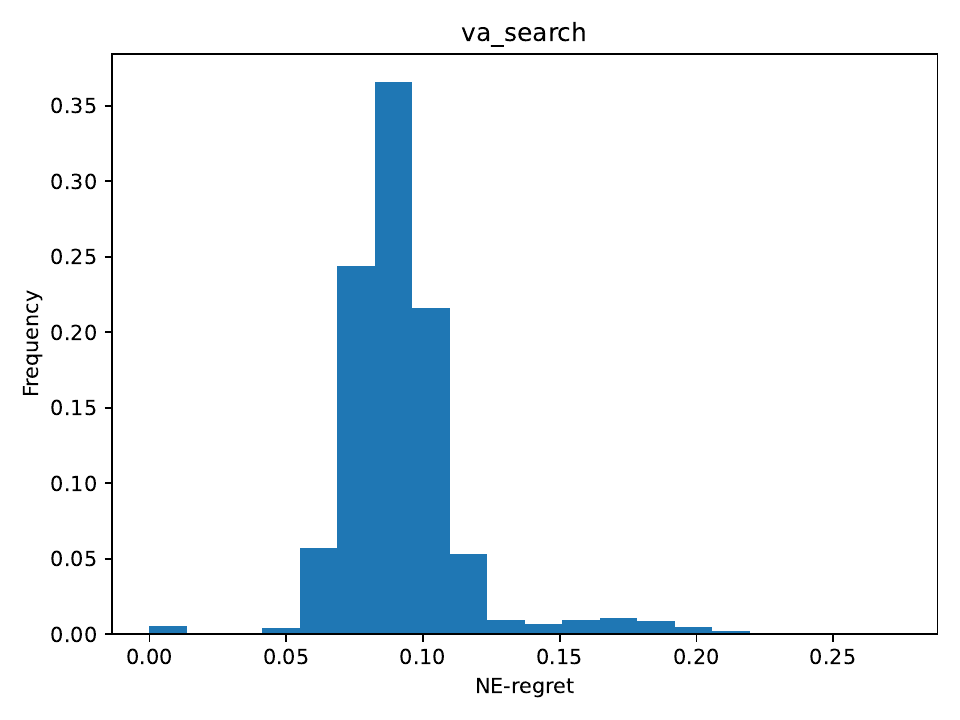} \\
\includegraphics[width=0.24\textwidth]{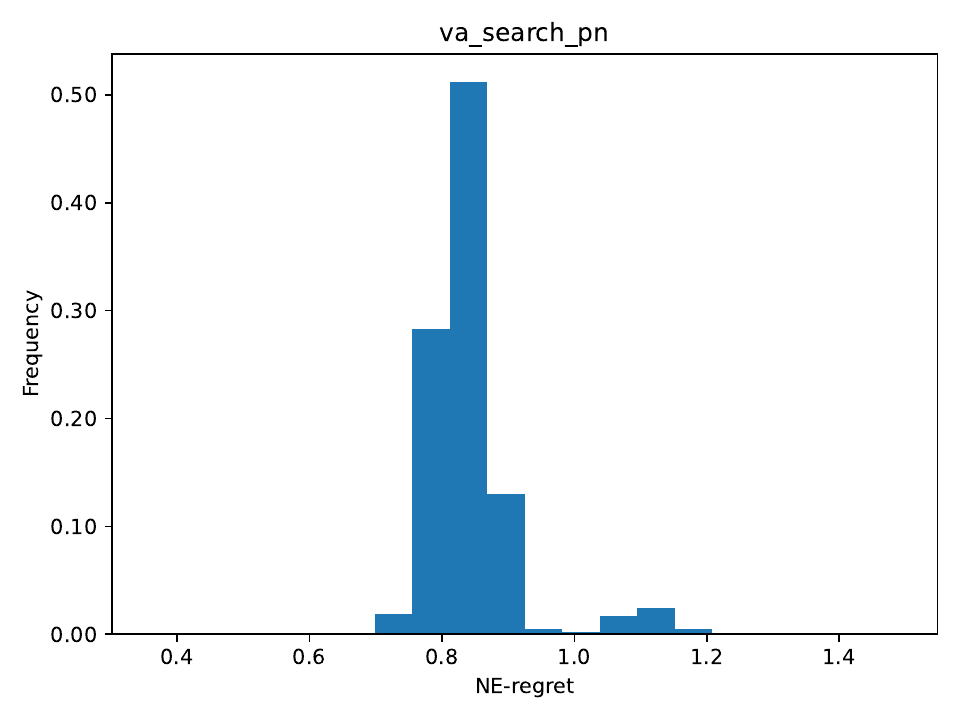}
\end{tabular}
\caption{Empirical Distribution of NE-Regret of $\Barg(10, 0, 1)$}
\label{tab:ne-regret-10}
\end{figure*}

See Figures~\ref{tab:ne-regret-10}.

\subsection{$\Barg(30, 0.125, 0.935)$}
\begin{figure*}[t]
\begin{tabular}{cccc}
\includegraphics[width=0.24\textwidth]{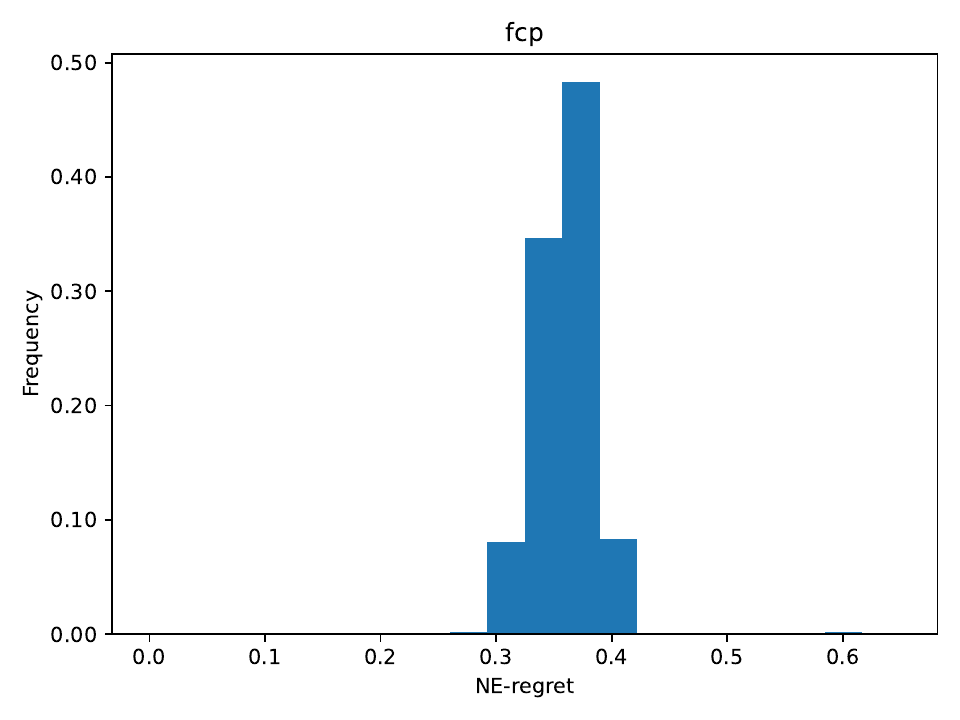} &
\includegraphics[width=0.24\textwidth]{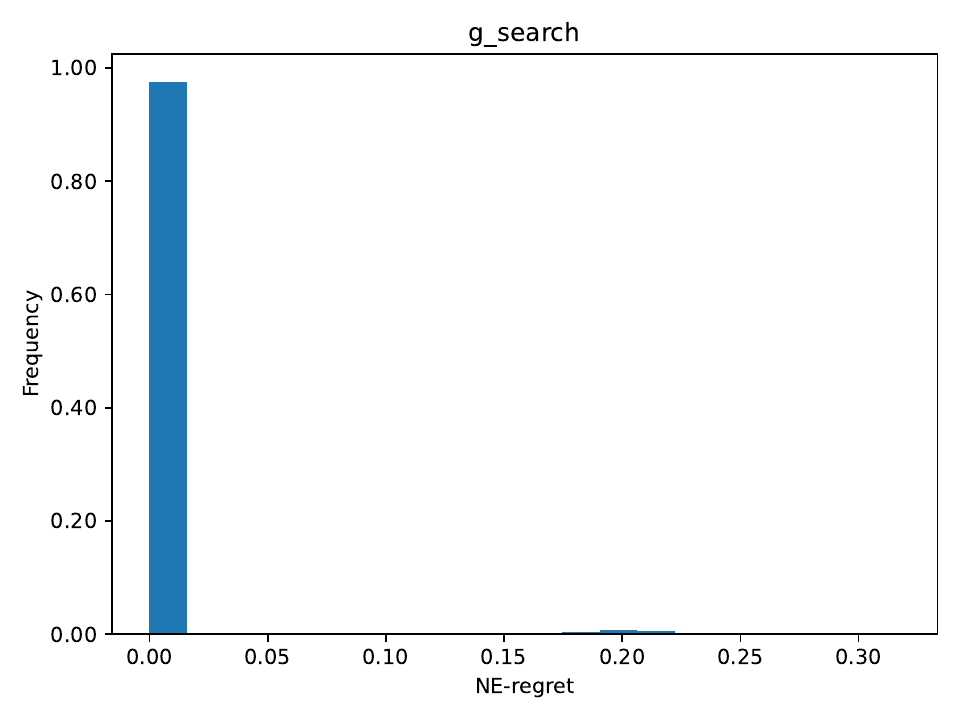} &
\includegraphics[width=0.24\textwidth]{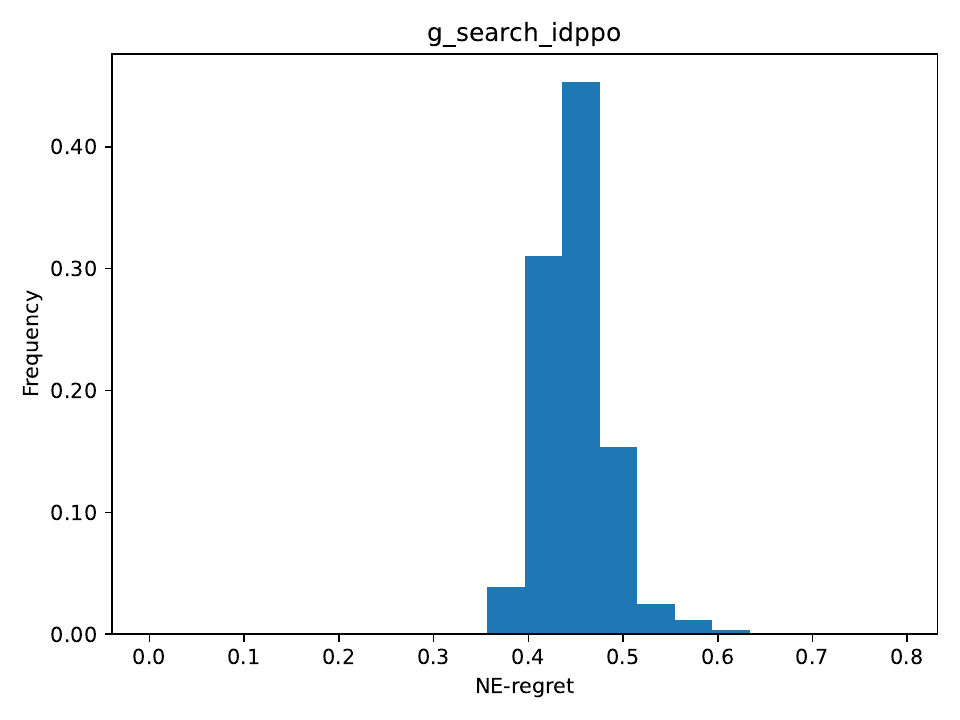} &
\includegraphics[width=0.24\textwidth]{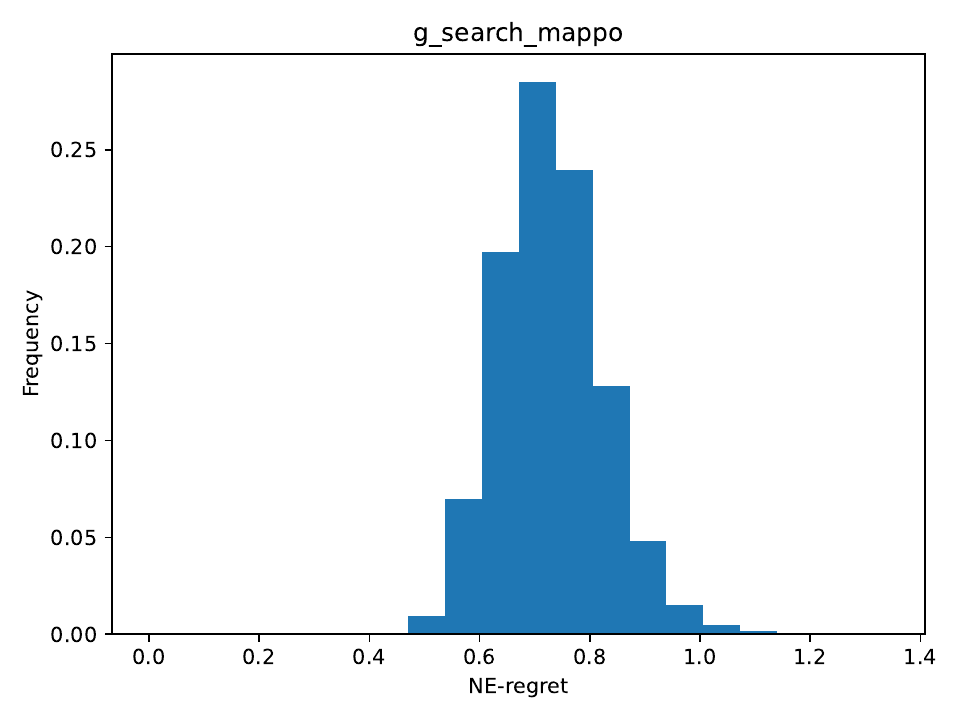} \\
\includegraphics[width=0.24\textwidth]{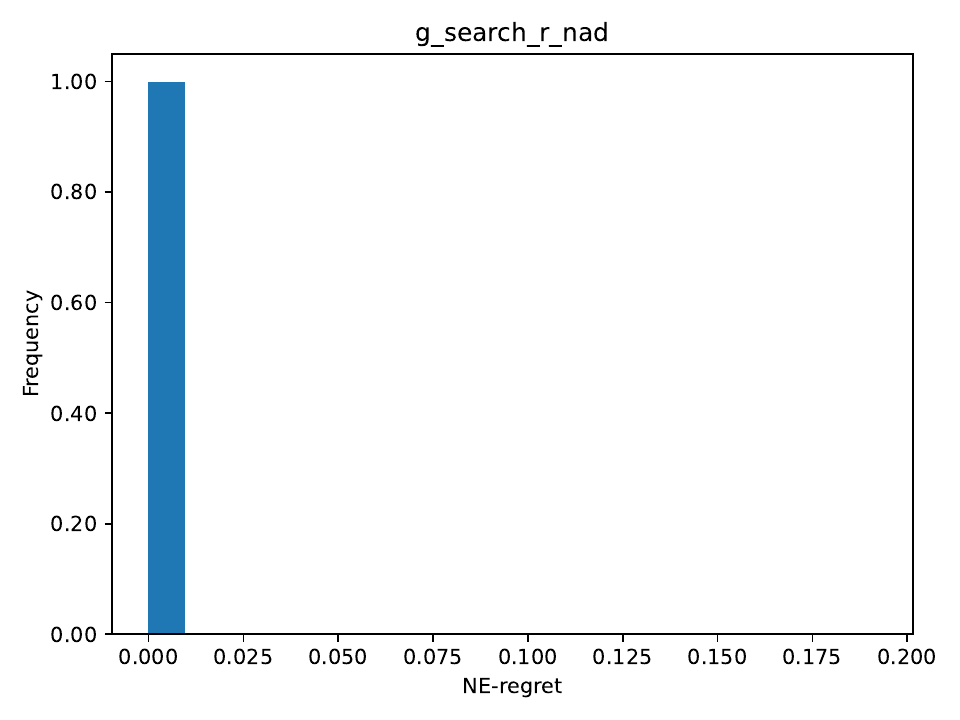} &
\includegraphics[width=0.24\textwidth]{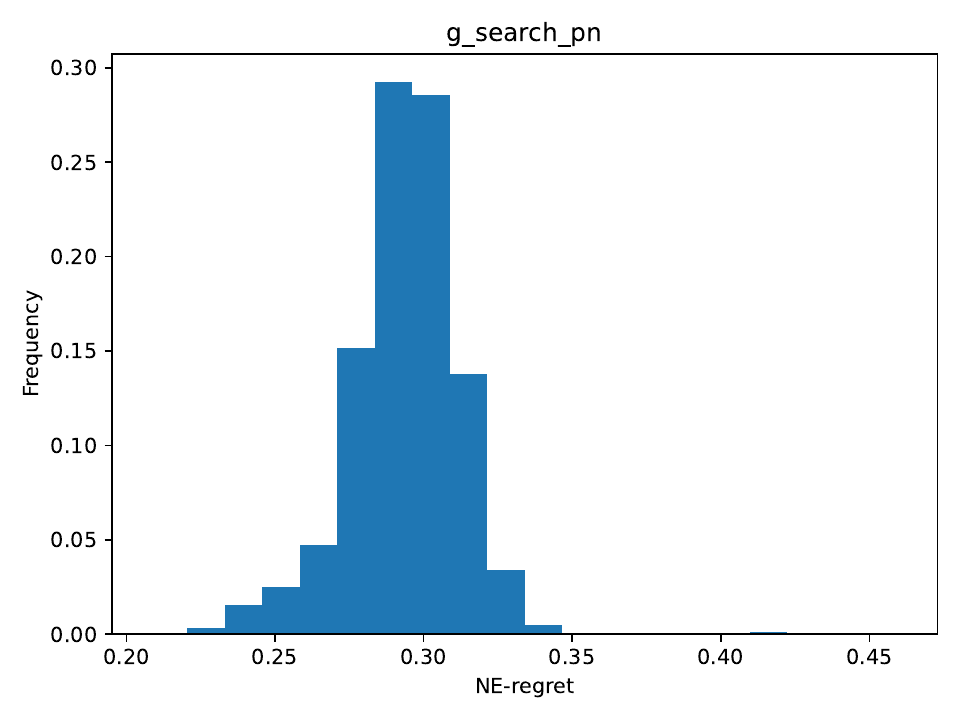} &
\includegraphics[width=0.24\textwidth]{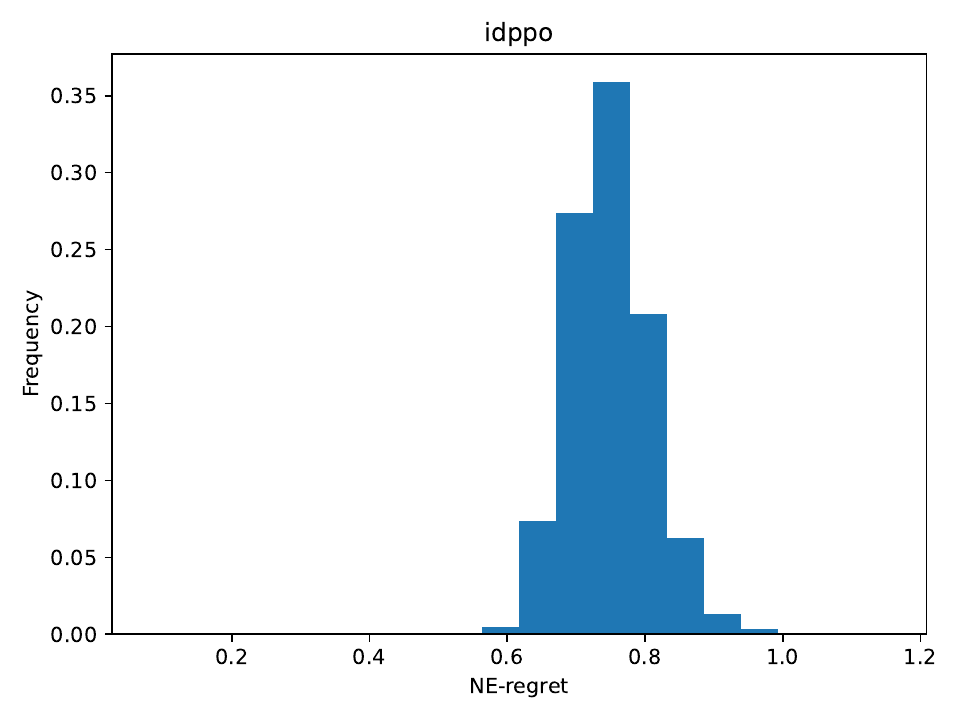} & 
\includegraphics[width=0.24\textwidth]{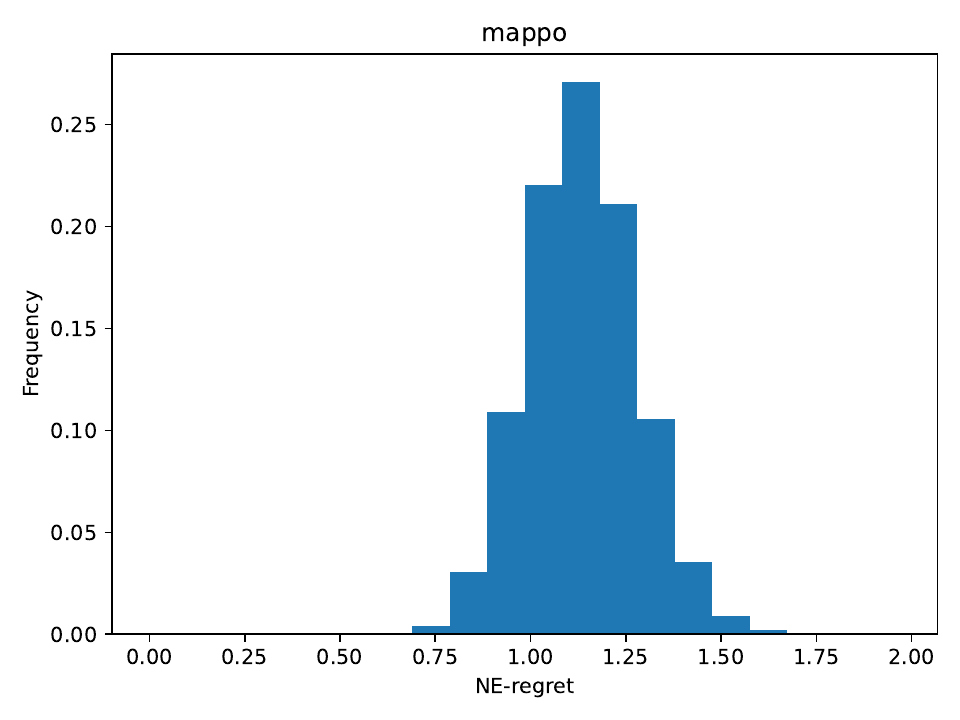}\\
\includegraphics[width=0.24\textwidth]{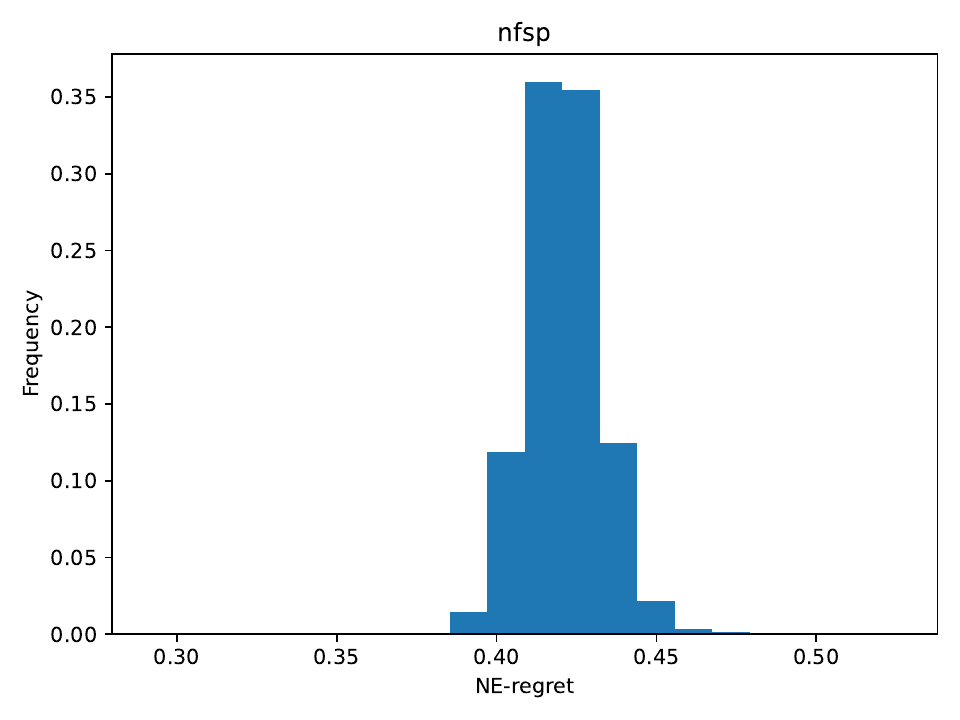} & 
\includegraphics[width=0.24\textwidth]{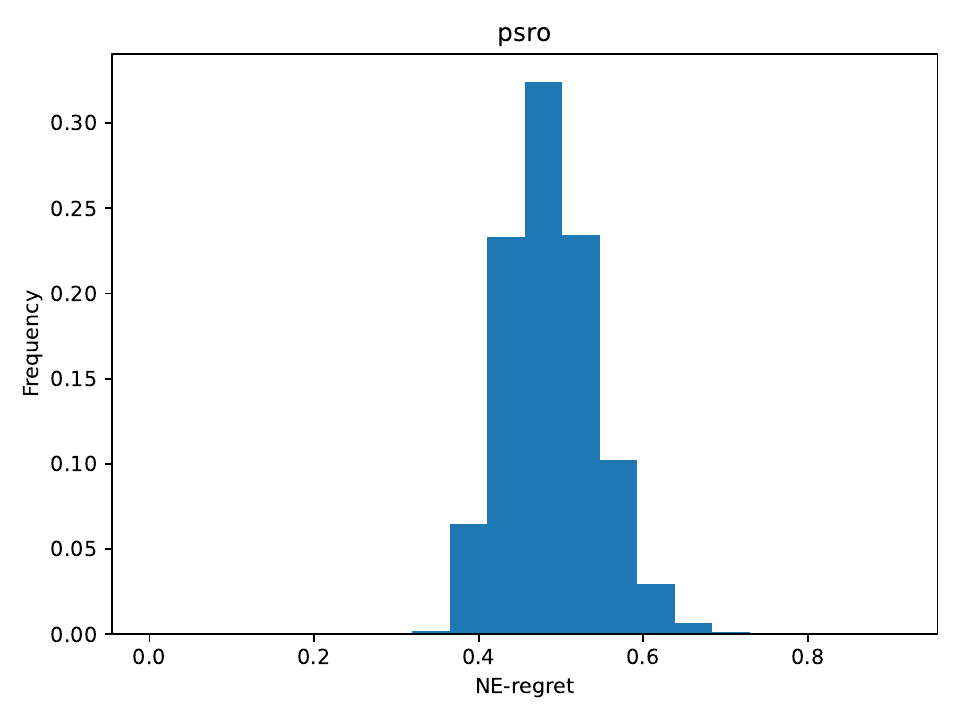} & 
\includegraphics[width=0.24\textwidth]{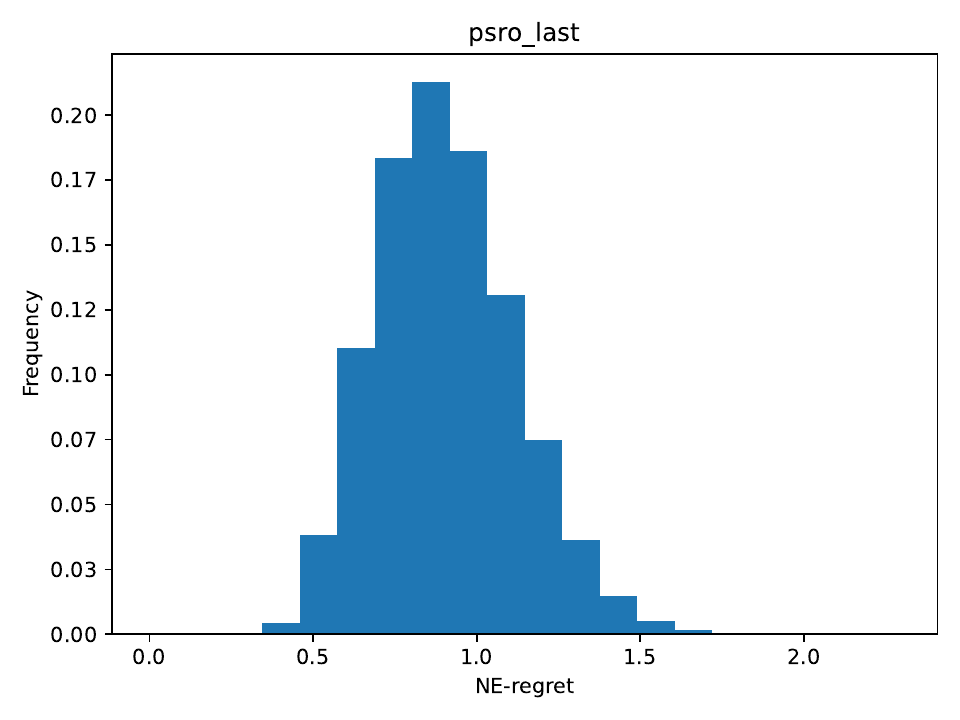} &
\includegraphics[width=0.24\textwidth]{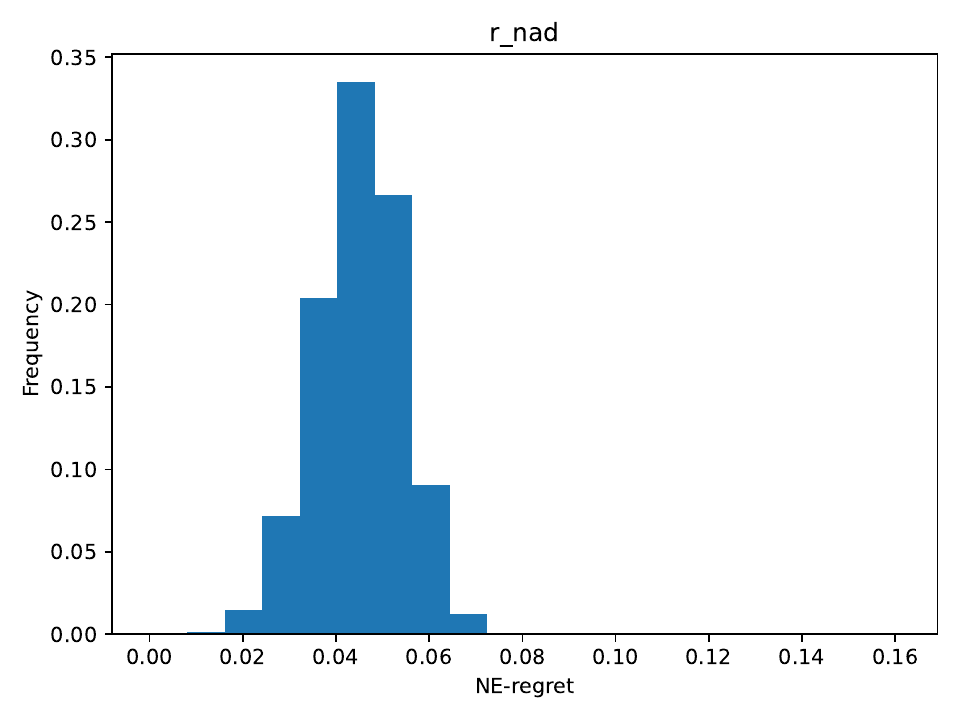}\\
\includegraphics[width=0.24\textwidth]{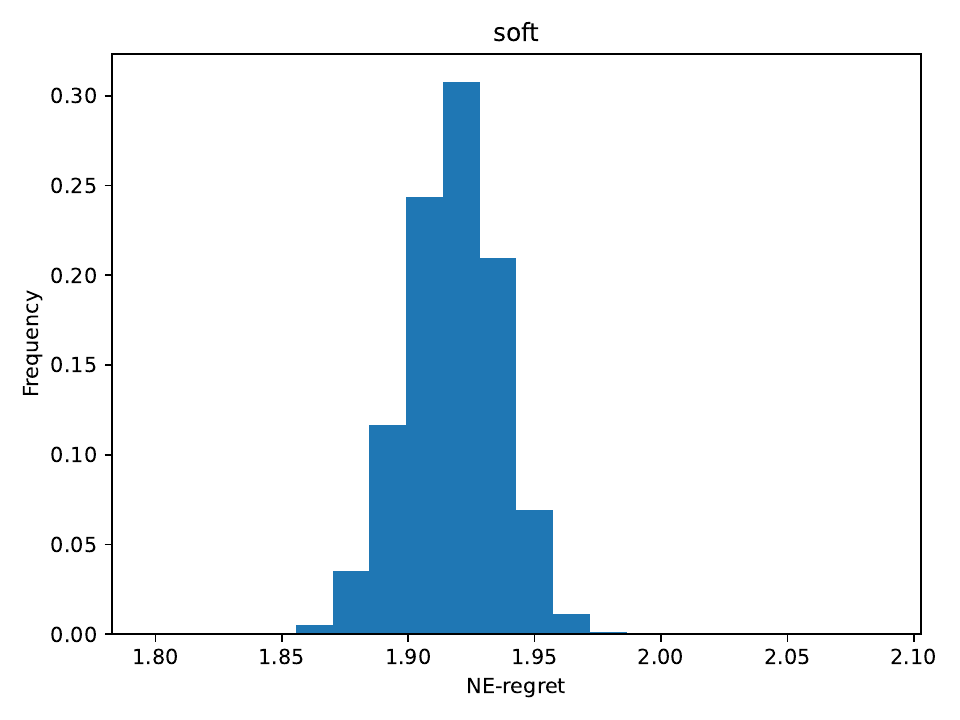} &
\includegraphics[width=0.24\textwidth]{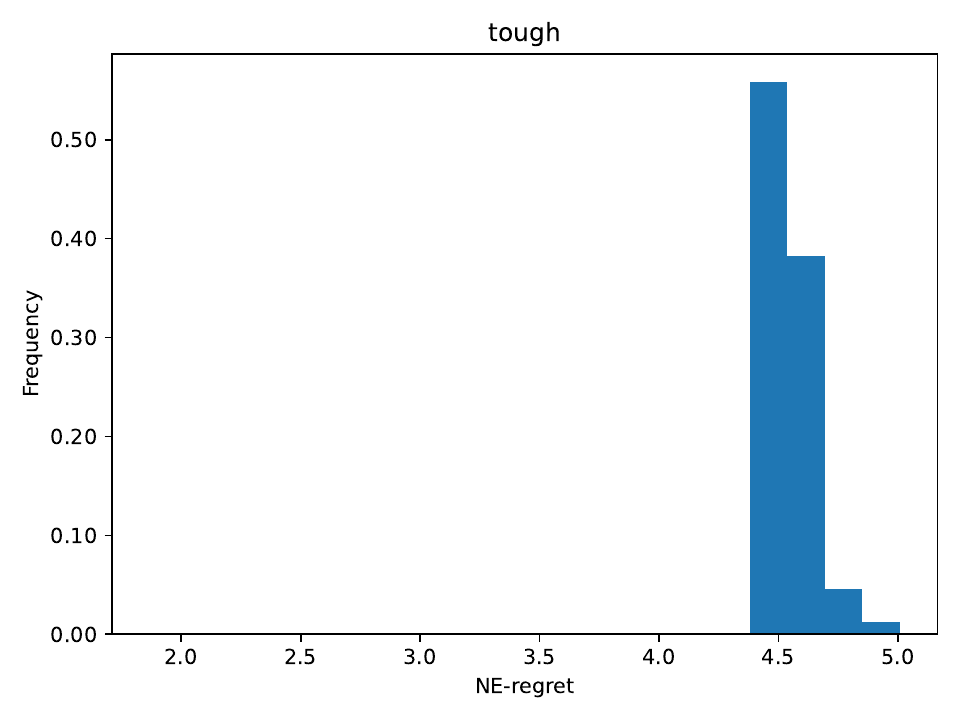} & 
\includegraphics[width=0.24\textwidth]{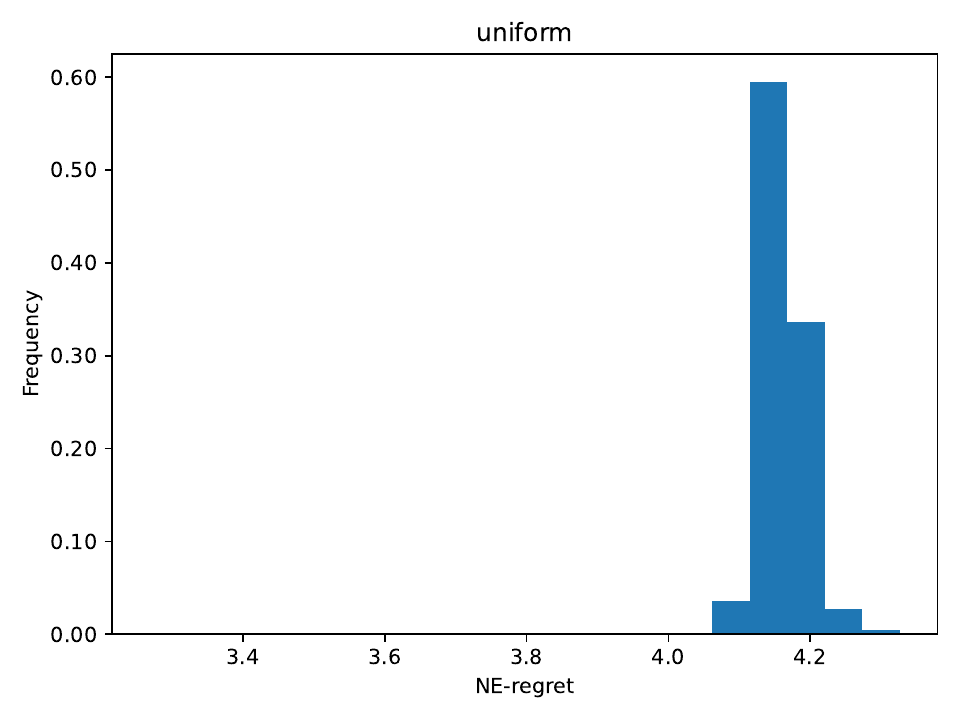} & 
\includegraphics[width=0.24\textwidth]{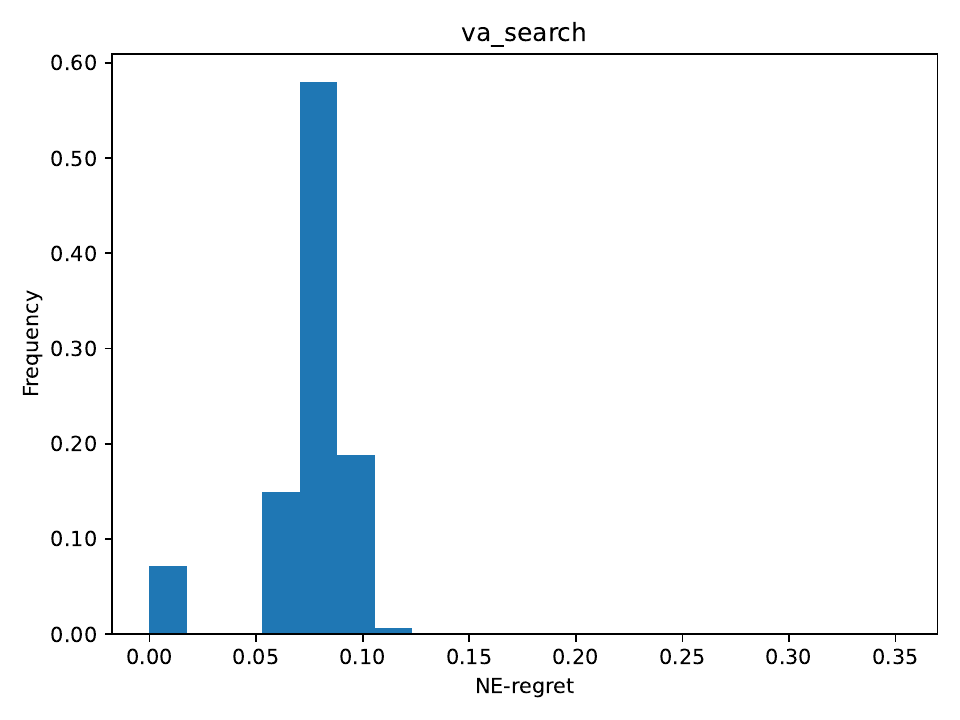} \\
\includegraphics[width=0.24\textwidth]{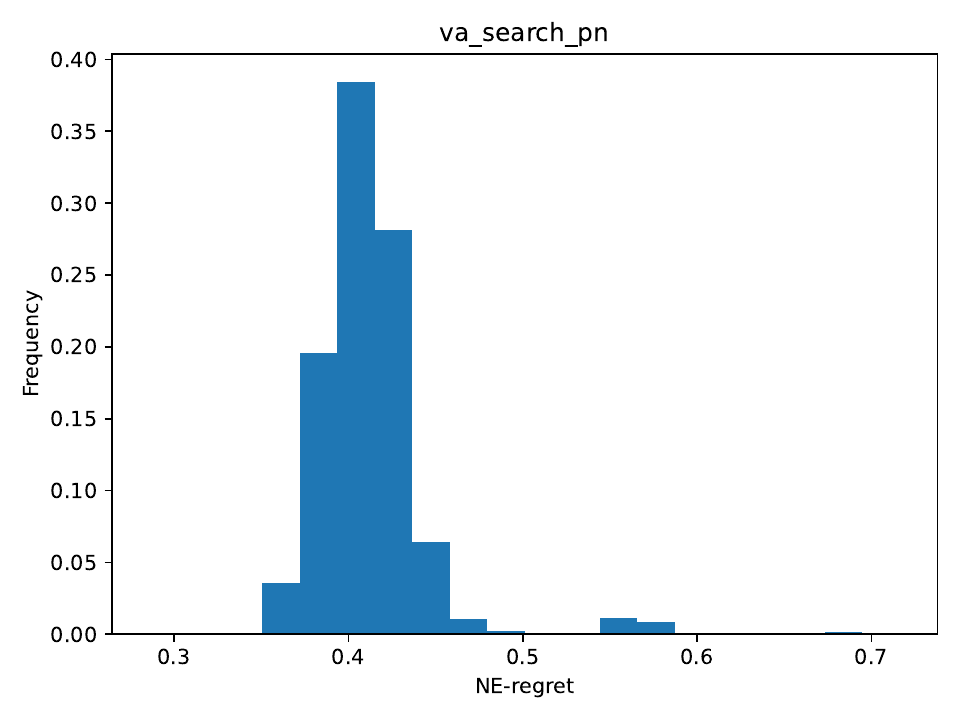}
\end{tabular}
\caption{Empirical Distribution of NE-Regret of $\Barg(30, 0.125, 0.935)$}
\label{tab:ne-regret-30}
\end{figure*}

See Figures~\ref{tab:ne-regret-30}

\section{Approximate Size of Deal or No Deal}
\label{sec:app-games-dond}

To estimate the size of Deal or No Deal, we first verified that there are 142 unique preference vectors per player. Then, we generated 10,000 simulations of trajectories using uniform random policy, computing the average branching factor (number of legal actions per player at each state) as $b \approx 23.5$. 

Since there are 142 different information states for player 1's first decision, about $142 b b$ player 1's second decision, etc. leading to $142 ( 1 + b^2 + b^4 + b^8) \approx 13.2 \times 10^{12}$ information states. Similarly, player 2 has roughly $142 (1 + b^1 + b^3 + b^5 + b^7) = 5.63 \times 10^{11}$ information states.

\end{document}